\documentclass[pra, floatfix, preprint, endfloats*]{revtex4}
\usepackage{textcomp}
\usepackage{soul}
\usepackage{slashed}
\usepackage[dvips]{graphicx}
\usepackage{mathrsfs}
\usepackage{amsfonts}
\usepackage{amsmath}
\usepackage{amssymb}
\usepackage{revsymb}
\usepackage{graphicx}
\usepackage{mathrsfs}
\usepackage{bm}
\usepackage{psfrag}

\newcommand{\be}{\begin{equation}}
\newcommand{\ee}{\end{equation}}
\newcommand{\ba}{\begin{array}}
\newcommand{\ea}{\end{array}}
\newcommand{\bea}{\begin{eqnarray}} 
\newcommand{\eea}{\end{eqnarray}} 
\newcommand{\bd}{\begin{displaymath}}
\newcommand{\ed}{\end{displaymath}}
\newcommand{\bc}{\begin{center}}
\newcommand{\ec}{\end{center}}
\newcommand{\units}[1]{\,\textrm{#1}}
\newcommand{\eps}{\varepsilon}
\newcommand{\mbf}[1]{\mathbf{#1}}
\newcommand{\mbfr}{\mathbf{r}}
\newcommand{\mbfrd}{\mathbf{r}_{\trm{d}}}
\newcommand{\trm}[1]{\textrm{#1}}
\newcommand{\rhv}{\mbf{\hat{r}}_{\trm{d}}}
\newcommand{\xrd}{\frac{x_{\trm{d}}}{r_{\trm{d}}}}
\newcommand{\yrd}{\frac{y_{\trm{d}}}{r_{\trm{d}}}}
\newcommand{\zrd}{\frac{z_{\trm{d}}}{r_{\trm{d}}}}

\newcommand{\matlab}{{\sc{matlab}} }
\newcommand{\gfunc}[2]{\frac{\exp\big[-i\omega\big(|\mbf{#1} - \mbf{#2}|+t\big)\big]}                                                  {|\mbf{#1} - \mbf{#2}|}}
\long\def\symbolfootnote[#1]#2{\begingroup%
\def\thefootnote{\fnsymbol{footnote}}\footnote[#1]{#2}\endgroup}
\newcommand{\figref}[1]{Fig. \ref{#1}}
\newcommand{\figrefa}[1]{Fig. \ref{#1}a}
\newcommand{\figrefb}[1]{Fig. \ref{#1}b}
\newcommand{\eqnref}[1]{Eq. (\ref{#1})}
\newcommand{\eqnrefs}[2]{Eqs. (\ref{#1}) and (\ref{#2})}
\newcommand{\sxnref}[1]{Sec. \ref{#1}}
\newcommand{\pref}[1]{P. \pageref{#1}}
\fussy
\def\crampest{\medmuskip = 1mu plus 1mu minus 1mu}
\def\uncramp{\medmuskip = 4mu plus 2mu minus 4mu}

\begin{document}
\title{Double-slit vacuum polarisation effects in ultra-intense laser fields}
\author{B. \surname{King}}
  \email{ben.king@mpi-hd.mpg.de}
\author{A. \surname{Di Piazza}}
  \email{dipiazza@mpi-hd.mpg.de}
\author{C. H. \surname{Keitel}}
  \email{keitel@mpi-hd.mpg.de}
  \affiliation{Max-Planck-Institut f\"ur Kernphysik,
    Saupfercheckweg 1, D-69117 Heidelberg, Germany}

\date{April 15, 2010}
\begin{abstract}
The influence of the strong laser-driven vacuum on a propagating electromagnetic probe wave has been studied in detail. We investigate two scenarios comprising a focused probe laser beam passing through a region of vacuum polarised by an ultra-intense laser field. By splitting this strong field into two, separated, monochromatic Gaussian pulses counter-propagating in a plane perpendicular to the probe field axis, we demonstrate a leading order light-by-light diffraction effect that generates an interference pattern reminiscent of the classic double-slit experiment. We calculate the total number of probe photons diffracted as well as the number diffracted into regions where the vacuum polarisation signal is higher than the probe background. In addition, we calculate the induced ellipticity and polarisation rotation in the probe beam and show how, in the realistic situation in which the centres of the two strong fields are not exactly aligned, certain ranges of beam separation and observation distance may actually lead to an increase over the idealised case of a single strong laser beam.
\end{abstract} 
\maketitle

\section{Introduction}
\setcounter{equation}{0}
\renewcommand{\theequation}{\Roman{section}.\arabic{equation}}
That strong electromagnetic fields can modify the dielectric properties of the quantum vacuum has been known since the pioneering work of Heisenberg and Euler, Weisskopf and Sauter \cite{heisenberg_euler36, weisskopf36, sauter31}. Quantum electrodynamics (QED) predicts that when the electromagnetic field strength nears the critical ``Schwinger limit'' required to spontaneously create an electron-positron pair with an electron of charge $-e<0$ and mass $m$ within the reduced Compton wavelength $\lambdabar_{c}=\hbar/mc$, a range of nonlinear vacuum polarisation effects (VPEs) should become observable. The corresponding electric field of $E_{\trm{cr}} = \sqrt{4\pi}m^2c^3/\hbar e = 1.3\times10^{16} \units{Vcm}^{-1} \!$ in Lorentz-Heaviside units, would certainly allow one to access \emph{attenuative} VPE processes, namely involving real electron-positron pair creation, whose rates would become large enough to be easily observed. Moreover, recent calculations show that these effects in the presence of loan fields can already be clearly observed at intensities orders of magnitude below critical values \cite{narozhny_creation06, kirk08, kirk09, dunne09}. Such VPE processes have also been discussed as a probe for new fundamental physics, with current limits clarified and new experiments proposed \cite{guendelman08, ahlers08, ahlers07, gies09, tommasini09}. Since these processes are exponentially suppressed, for the case in earth-bound laboratories where the electric fields involved are much less than $E_{\trm{cr}}$, it is \emph{refractive} VPE processes involving virtual electron-positron pairs that are most likely to be observed. Photon-photon scattering is one example of a refractive VPE \cite{dittrich85}, which has already been carried out as Delbr\"uck scattering involving virtual photons in the Coulomb field of a heavy nucleus \cite{milstein94} where the atomic number $Z$ is $\lesssim1/\alpha$ and $\alpha = e^2/4 \pi \hbar c\approx1/137$ is the fine-structure constant, but has since eluded detection for purely real photons \cite{bernard00}. This effect could be measured by virtue of polarisation-dependent emission in four-wave mixing \cite{lundstroem_PRL_06}; by using the transverse-electric modes of plane waves to generate a resonant coupling in a waveguide \cite{marklund_PRL_01} or using sufficiently intense lasers to compensate for the small cross-section and inducing a phase-shift in lasers passing through one another \cite{tommasini08, Ferrando_scattering07} (for a review of the applications of relativistic lasers see \cite{salamin_review06, mourou_review06, marklund_review06}). Other nonlinear vacuum effects include ``photon acceleration'' \cite{mendonca06}; photon splitting in atomic fields \cite{akhmadaliev02}, pair plasmas \cite{Brodin_splitting_PRL_01} and laser fields \cite{dipiazza_splitting07}; as well as the corresponding reverse process of vacuum high-order harmonic generation e.g. in various laser set-ups \cite{dipiazza_harmonic05, fedotov_harmonics06} or in a mixed Coulomb and laser field set-up \cite{jentschura_harmonics05}. The corresponding critical bound of the magnetic field, $B_{\trm{cr}} = \sqrt{4\pi}m^2c^3/\hbar e = 4.4\times10^{13}\units{G}$ can be surpassed by ultramagnetised neutron stars or ``magnetars'' \cite{ozel_xraypulsars07} (for a review on X-ray pulsars, see \cite{kaspi_pulsars07}), which provide an inhomogeneous trigger for nonlinear effects such as vacuum birefringence and photon ray-bending \cite{denisov_pulsar03} as well as photon-splitting \cite{baring08, Usov_pulsar02}. The possibility of laboratory-based experiments that measure second-harmonic generation within a constant inhomogeneous magnetic field, have also been considered \cite{ding91}. The current PVLAS (Polarizzazione del Vuoto con Laser) experiment uses a slowly-varying magnetic field to attempt to detect refractive-regime vacuum-induced birefringence and dichroism through rotation in the polarisation of a probe laser wave \cite{pvlas08a}. In addition to in a magnetic field, birefringence can also be induced in the vacuum by e.g. a laser field in this regime \cite{king10a, marklund10, heinzl_birefringence06, dipiazza_PRL_06}. This latter scenario, and that of vacuum-induced diffraction are two examples of refractive VPEs which we further develop in the current paper.
\newline


At the time of writing, the record for the highest intensity laser ever produced is held by the {\sc hercules} laser and stands at $2 \times 10^{22}\units{Wcm}^{-2}$ \cite{yanovsky08}, seven orders of magnitude removed from the Schwinger limit intensity of $I_{\trm{cr}} = cE^{2}_{\trm{cr}} / 2 = 2.3 \times 10^{29}\units{Wcm}^{-2} \! $. We foresee that with the next generation of lasers currently being built, we will soon be in a much better position to test vacuum effects and so work with the quoted values for intended intensity and photon energy ranges in the coming decade. Examples of strong-field lasers are the ELI (Extreme Light Infrastructure) and HiPER (High Power laser Energy Research) facilities with target intensity values of 10$^{26}$ W\trm{cm}$^{-2}$ \cite{ELI_SDR, HiPER_TDR}. The PFS (Petawatt Field Synthesiser, \cite{karsch09}), whilst planning a lower intensity of 10$^{22} \units{Wcm}^{-2}$, will have a repetition rate of 10$\units{Hz}$ which could be more favourable in certain situations and provides an example of a cutting edge system to be commissioned later this year. Free-Electron Lasers (FELs) where undulating electrons provide the lase medium, such as the XFEL (X-Ray Free-Electron Laser) and the LCLS (Linac Coherent Light Source) could also be used to polarise the quantum vacuum, especially with the so-called ``goal'' parameters quoted in \cite{ringwald01}. However, a further application of the FELs, one which could be reached sooner, would be as probe field lasers, whose alteration when passing through vacuum polarised regions could be measured. The XFEL and LCLS would be ideal for measuring refractive effects, which are in general proportional to laser frequency, as they allow continuous adjustment of the probe wavelength down to a minimum of 0.1\units{nm} and 0.15\units{nm} respectively \cite{XFEL_TDR, LCLS}.
\newline

This paper concerns itself with laser-induced vacuum-polarisation effects in the spirit of \cite{dipiazza_PRL_06}. Here, the change in polarisation and ellipticity of a planar Gaussian probe field passing through a region of the vacuum polarised by a perpendicular standing wave formed by two counter-propagating, ultra-intense ($I_{0} \geq 10^{23}\units{Wcm}^{-2}$) Gaussian beams, was calculated up until the point where probe defocusing becomes important. We compare and expand upon this simple set-up with the following enhancements:
\renewcommand{\labelenumi}{\roman{enumi}.}
\begin{enumerate}
\item The two counter-propagating strong field wave triggers for VPEs are separated in the plane perpendicular to their propagation, modelling a more realistic situation. This makes sense first from an experimental point of view, to know how VPEs are sensitive to laser alignment, and second allows us to derive an interference effect as different parts of the probe beam pass through the ``double-slit''-like, vacuum-polarised region. This will even turn out to increase probe-beam polarisation rotation and ellipticity.
\item Defocusing terms were introduced into the probe beam and a corresponding update to the vacuum-induced ellipticity and rotation of the probe polarisation. This extends the limited range of detector distances in \cite{dipiazza_PRL_06} where we could have compared theory to experiment, as our new expressions are also valid in the far-field diffraction zone, where they converge to a non-zero value.
\item The electric field generated in the new set-up by the current of the polarised vacuum, which we henceforth label the ``diffracted field,''  was also calculated in the probe beam's transverse plane, allowing us to again model the more realistic situation where a detector is placed off-axis, in regions where the diffracted field, having a wider spread, is larger than the probe background.
\end{enumerate} 
Throughout, we will make the analogy with the single- and double- slit diffraction experiment. According to Babinet's principle, the diffraction pattern generated by light passing an opaque obstacle is the same as that for light traversing an aperture with the same shape as the obstacle \cite{levi68}. Regions of the vacuum polarised by the two strong-field laser beams then represent ``translucent'' obstacles for photons in the probe beam, having as we will show, a non-trivial polarisation and magnetisation. Unlike the typically sharp two-dimensional slits used in demonstration experiments, the strong lasers, being Gaussian in beam profile, form smooth, three-dimensional slits. One consequence of this will be that no single-slit fringes occur in the far field. However, the probe photons scattered from each strong beam will interfere with one another, and in this way, we \emph{will} have a ``double-slit''. As the scattering of probe photons occurs with such a small probability, the complete double-slit pattern will only be observable when the background of probe photons passing unperturbed through the apparatus, is subtracted. At the detector, the total field will in general consist of the probe signal plus the vacuum contribution of the scattered photons. In calculating the interference between these two fields, we demonstrate a new diffractive effect of the polarised vacuum, which we accentuate by forming the double-slit-like experimental set-up. In addition, we also compare polarisation results with a second beam geometry, namely that of the probe propagating anti-parallel to the strong field, which we label the ``double-shaft'' (or ``single-shaft'')  set-up. Our results are complementary to findings in \cite{king10a}, which focus on only the pure diffracted intensity for a different laser geometry. 
\newline

The paper is organised as follows: in section II we first introduce the Euler-Heisenberg theory upon which the results are based and the range of experimental scenarios we consider; then follows in section III an analysis of the first part of the results, the intensity of the bare diffracted field and the time-averaged difference in `probe + diffracted' signal in terms of number of photons; the second part of the results deals with the change in rotation and ellipticity of polarisation for both of the two geometries and the paper is concluded with a recapitulation of the main results.
%
%
\section{Theoretical basis}
\setcounter{equation}{0}
\renewcommand{\theequation}{\Roman{section}.\arabic{equation}}
\subsection{Leading-order vacuum current}
\setcounter{equation}{0}
\renewcommand{\theequation}{\Roman{section}.\arabic{equation}}
By making two basic assumptions, we can drastically simplify the interaction terms occurring in our field theory \cite{gies00}. From the assumption that the photon energies involved are much lower than the electron rest energy, follows that the loop contribution of spatio-temporal extent can be consistently regarded as a single local interaction point, thereby allowing us to write down the so-called Euler-Heisenberg local, point Lagrangian density, $\mathcal{L}$, which we will use to describe vacuum polarisation effects. Secondly, from the aforementioned laser intensities which are either currently available or scheduled for the future, we will work with the comfortable assumption that field strengths are much lower than critical values. This then allows us to use the weak-field expansion of the Euler-Heisenberg Lagrangian, the leading order of which (in a system of units adopted henceforth, $\hbar=c=1$) reads:
\be \label{eqn:eh_lagrangian}
\mathcal{L} = \frac{1}{2} (E^{2} - B^{2}) + \frac{2\alpha^{2}}{45m^{4}} 
              \big[ (E^{2} - B^{2})^{2} + 7(\mathbf{E} \cdot \mathbf{B})^{2} \big],
\ee
for electric and magnetic fields $\mbf{E}$ and $\mbf{B}$ and their square moduli $E^{2} = \mbf{E} \cdot \mbf{E}$,  $B^{2} = \mbf{B} \cdot \mbf{B}$ respectively. Extremising the action with respect to the vector potential corresponding to these fields, we achieve the following wave equations of motion, for an induced \emph{vacuum current} $\mbf{J}_{\trm{vac}}$:
\bea \label{eqn:eom}
\nabla^{2} \mbf{E} - \partial_{t}^{2} \mbf{E} & = & \mbf{J}_{\trm{vac}} = \nabla \wedge \partial_{t}\mbf{M}
                                                     -\nabla (\nabla \cdot \mbf{P})
                                                     +\partial_{t}^{2}\mbf{P}, \\
\quad \label{eqn:pol} \mbf{P} & := & \phantom{-}\frac{4 \alpha^{2}}{45 m^4} \big[ 2(E^{2} - B^{2}) \mbf{E} 
                                             + 7(\mbf{E} \cdot \mbf{B}) \mbf{B} \big], \\
\label{eqn:mag} \mbf{M} & := & -\frac{4 \alpha^{2}}{45 m^4} \big[ 2(E^{2} - B^{2}) \mbf{B} 
                                          - 7(\mbf{E} \cdot \mbf{B}) \mbf{E} \big].
\eea
There are many similarities one can draw between birefringent solid-state materials and the behaviour of the vacuum under intense electromagnetic fields. Direct from the above wave equation \eqnref{eqn:eom}, we can liken the vacuum current to one representing the response from such a birefringent material, that is to say, labelling \textbf{P} its polarisation, and \textbf{M} its magnetisation \footnote{Due to a printing error on page 2 of \cite{dipiazza_PRL_06}, the magnetisation appears first here with the correct overall minus sign.}. As these are functions of both $\mbf{E}$ and $\mbf{B}$, the inhomogeneity in our fields which here plays a central role, is included at this point.

\subsection*{Definition of experimental scenario}
\label{sxn:defscen}
\begin{figure}[!hp]
\noindent\centering
\includegraphics[width=4in]{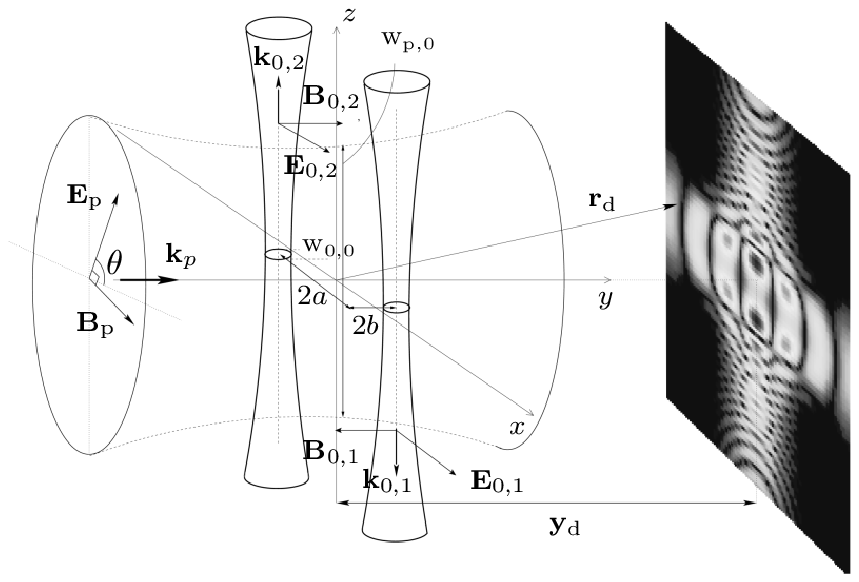}
\caption{The double-slit experimental setup. A monochromatic Gaussian probe beam with electric and magnetic field vectors $\mbf{E}_{\trm{p}}$ and $\mbf{B}_{\trm{p}}$ respectively, wavevector $\mbf{k}_{\trm{p}}$, linearly polarised at an angle $\theta$ to the $x$-axis in the $x$--$z$ plane and with a waist $\trm{w}_{\trm{p},0}$, much greater than the strong-field beam waist, $\trm{w}_{0,0}$, impinges and is perpendicular to two, parallel, counter-propagating, monochromatic and Gaussian strong-field waves with amplitudes $E_{0}/\sqrt{2} \gg E_{\trm{p}}$, electric and magnetic fields in the $x$-$y$ plane $\mbf{E}_{0,1}, \mbf{E}_{0,2}$ and $\mbf{B}_{0,1}, \mbf{B}_{0,2}$ respectively, wavevectors $\pm\mbf{k}_{0} = (0,0,\mp{\omega}_{0})$, $\omega_{0} \ll \omega_{\trm{p}}$, with foci at $(x,y) = (a,b)$ and $(x,y) = (-a,-b)$. The results of this process are then measured a distance $r_{\trm{d}}$ from the centre of the interaction region.\label{fig:exp_setup}} 
\end{figure}

In this paper, we will focus mainly on the double-slit set-up sketched in \figref{fig:exp_setup}, and include only a summary of the polarisation results for the ``single-shaft'' set-up corresponding to a head-on probe and strong-field collision, towards the end of the article (we use here the label ``shaft'' contrary to in \cite{king10a}, where it was also labelled a ``slit''). For our double-slit set-up, two tightly-focused (we assume the diffraction limit has been reached i.e. focused down to a wavelength), counter-propagating monochromatic strong field Gaussian pulses polarising the vacuum, with waists $\trm{w}_{0,0}$ centred at $(x,y)=(a,b)$ and $(x,y)=(-a, -b)$, electric fields $\mbf{E}_{0,1}(x,y,z,t)$, $\mbf{E}_{0,2}(x,y,z,t)$ and wavevectors $\mbf{k}_{0}=(0,0,-\omega_{0})$ and $-\mbf{k}_{0}=(0,0,\omega_{0})$ respectively, are permeated simultaneously by a broader and weaker linearly-polarised, transverse monochromatic Gaussian probe field, polarised at an angle $\theta$ to the x-axis with waist $\trm{w}_{\trm{p},0}$, electric field $\mbf{E}_{\trm{p}}$ and wavevector $\mbf{k}_{\trm{p}}=(0,\omega_{\trm{p}},0)$, which will gain a diffracted component, whose intensity and polarisation will be measured some distance, $r_{\trm{d}}$ away. Using the Gaussian beam solution from \cite{salamin_review06}, in the effective interaction region we therefore have the following:
\bea 
\mbf{E}_{0}(x,y,z,t)\!\! & := & \!\!\Big[E_{0,1}(x,y,z,t) + E_{0,2}(x,y,z,t)\Big]\mbf{\hat{x}}, \label{eqn:E_0} \\[1.5ex]
E_{0,1}(x,y,z,t)\!\! & := & \!\!E_{0,0}(x\!-\!a,y\!-\!b,z) \sin\!\!\Big(\!\psi_{0} + \omega_{0}t + \omega_{0}z - 	\phi_{\trm{g},0}({z}) + \frac{\omega_{0}z}{2} \frac{(x-a)^{2} + (y-b)^{2}}{z^{2} + z_{\trm{r},0}^{2}}\Big),\nonumber\\
E_{0,2}(x,y,z,t)\!\! & := & \!\!E_{0,0}(x\!+\!a,y\!+\!b,z) \sin\!\!\Big(\!\psi_{0} + \omega_{0}t - \omega_{0}z + 	\phi_{\trm{g},0}({z}) - \frac{\omega_{0}z}{2} \frac{(x+a)^{2} + (y+b)^{2}}{z^{2} + z_{\trm{r},0}^{2}}\Big),\nonumber\\
\mbf{E}_{\trm{p}}(x,y,z,t)\!\! & := &\!\! E_{\trm{p},0}(x, y, z) \sin\!\!\Big(\!\psi_{\trm{p}} + \omega_{\trm{p}}t - \omega_{\trm{p}}y 
                     + \phi_{\trm{g},\trm{p}}({y}) - \frac{\omega_{\trm{p}}y}{2} \frac{x^{2} + z^{2}}{y^{2}+y_{\trm{r},\trm{p}}^{2}}\Big) (\mbf{\hat{x}}\cos\theta\!+\!\mbf{\hat{z}}\sin\theta) \nonumber,\\
\eea
where we have defined respectively the strong and probe fields amplitudes $E_{0,0}(x,y,z)$, $E_{\trm{p},0}(x,y,z)$ with their maximum values $E_{0}/\sqrt{2}$ and $E_{\trm{p}} \ll E_{0}$, as:
\bea
E_{0,0}(x,y,z) := \frac{E_{0}}{\sqrt{2}} 
                          \frac{\mbox{e}^{-(x^{2} + y^{2})/\trm{w}_{0}^{2}}} 
                                          {\sqrt{1+(z/z_{\trm{r},0})^{2}}},
\qquad E_{\trm{p},0}(x,y,z) := 
	E_{\trm{p}}\frac{\mbox{e}^{-(x^{2} + z^{2})/\trm{w}_{\trm{p}}^{2}}}{\sqrt{1+(y/y_{\trm{r},\trm{p}})^{2}}}. \nonumber
\eea
The square of the waist of focusing is defined from beam parameters as $\trm{w}_{0}^{2} := \trm{w}^{2}_{0,0}(1+(z/z_{\trm{r},0})^{2})$, $\trm{w}_{\trm{p}}^{2} := \trm{w}^{2}_{\trm{p},0}(1+(y/y_{\trm{r},\trm{p}})^{2})$, where $\trm{w}_{\trm{p},0} \gg \trm{w}_{0,0}$, with Rayleigh lengths defined in the usual way, $z_{\trm{r},0} = \omega_{0}\trm{w}_{0,0}^{2}/2=\pi \trm{w}_{0,0}$ (as we have assumed $\trm{w}_{0,0}=\lambda_{0}$), $y_{\trm{r},\trm{p}} = \omega_{\trm{p}}\trm{w}_{\trm{p},0}^{2}/2$, and the Gouy phases respectively $\phi_{\trm{g},0}(z) = \tan^{-1}(z/z_{\trm{r},0})$, $\phi_{\trm{g},\trm{p}}(y) = \tan^{-1}(y/y_{\trm{r},\trm{p}})$. The fields in \eqnref{eqn:E_0} are chosen as a first-order approximation to the solution of Maxwell's equations in vacuum (see e.g. \cite{salamin_review06}, p.p. 64--65 for details on the higher order terms in this expansion), being an expansion in the small parameters $\epsilon_{z} = \trm{w}_{0,0}/z_{\trm{r},0} = \lambda_{0}/(\pi \trm{w}_{0,0}) \approx 1/\pi$ and $\epsilon_y = \trm{w}_{\trm{p},0}/y_{\trm{r},\trm{p}} = \lambda_{\trm{p}}/(\pi \trm{w}_{\trm{p},0}) \ll 1$ (as by definition the probe is not intensely focused and so $\trm{w}_{\trm{p},0} \gg \lambda_{\trm{p}}$). Therefore, throughout this calculation, we are working to an accuracy given by the largest term neglected in the expansion, $\epsilon_{z}$.
\newline

The magnetic fields consistent with this level of approximation are then:
\bea
\mbf{B}_{0}(x,y,z,t) & = & -\big[E_{0,1}(x,y,z,t) - E_{0,2}(x,y,z,t)\big]\hat{\mbf{y}},\\
\mbf{B}_{\trm{p}}(x,y,z,t) & = & E_{\trm{p},0}(x,y,z)(\hat{\mbf{x}}\sin\theta - \hat{\mbf{z}}\cos\theta) \times\nonumber\\[-0.5ex]
		& & \qquad \sin\Big(\psi_{\trm{p}} + \omega_{\trm{p}}t - \omega_{\trm{p}}y 
                        + \phi_{\trm{g},\trm{p}}({y}) - \frac{\omega_{\trm{p}}y}{2} \frac{x^{2} + z^{2}}{y^{2}+y_{\trm{r},\trm{p}}^{2}}\Big).
\eea

Since the probe field's strength is much lower than the strong field's, we regard terms $\propto E_{\trm{p}}^{2}, B_{\trm{p}}^{2}$ and smaller as being negligible in \eqnrefs{eqn:pol}{eqn:mag}. In addition, as we are only interested in the effects on the probe field, we drop terms which only depend on the strong field $\propto E_{0}^{3}, B_{0}^{3}$. With $\mbf{E} = \mbf{E}_{0} + \mbf{E}_{\trm{p}}$ and $\mbf{B} = \mbf{B}_{0} + \mbf{B}_{\trm{p}}$, the vacuum polarisation and magnetisation then becomes:
\bea 
\mbf{P} & = & \phantom{-}\frac{4 \alpha^{2}}{45m^{4}} \Big[ 2(  \mbf{E}_{0}\cdot\mbf{E}_{0} - 
								\mbf{B}_{0}\cdot\mbf{B}_{0}) \mbf{E}_{\trm{p}} 
				 + 4(\mbf{E}_{0} \cdot \mbf{E}_{\trm{p}}) \mbf{E}_{0}
				 + 7(\mbf{E}_{0} \cdot \mbf{B}_{\trm{p}}) \mbf{B}_{0} \Big], \label{eqn:pol_2} \\
\mbf{M} & = & -\frac{4 \alpha^{2}}{45m^{4}} \Big[ 2(    \mbf{E}_{0}\cdot\mbf{E}_{0} - 
							\mbf{B}_{0}\cdot\mbf{B}_{0}) \mbf{B}_{\trm{p}} 
				 + 4(\mbf{E}_{0} \cdot \mbf{E}_{\trm{p}}) \mbf{B}_{0}
				 - 7(\mbf{E}_{0} \cdot \mbf{B}_{\trm{p}}) \mbf{E}_{0} \Big]. \label{eqn:mag_2}
\eea
\subsection*{Diffracted field off-axis}
We will focus on the diffracted electric field $\mbf{E}_{\trm{d}}(\mbfrd, t)$, generated by the polarised vacuum current in \eqnref{eqn:eom}, at a displacement $\mbfrd$, from the centre of the interaction volume (centred at the origin of the co-ordinate system), in the direction of propagation of probe beam. 
Using Green's functions to solve the inhomogeneous wave equation driven by a current $\mbf{J}(\mbf{r}, t)$, we have, in general, $\protect{\mbf{E}_{\trm{d}}(\mbf{r}_{\trm{d}}, t) = -1/(4\pi) \int d^{3}r |\mbfrd - \mbfr|^{-1}\mbf{J}(\mbfr, t- |\mbfrd - \mbfr|)}$ \cite{jackson75}. It can be seen from the definition of our current in \eqnref{eqn:eom}, that we are going to have terms cubic in the electromagnetic field, which means cross-terms between our probe and strong fields in the interaction region. As our waves are monochromatic, we see that photons of discrete energies $\omega_{\trm{p}}$ and $\omega_{\trm{p}} \pm 2\omega_{0}$, etc. will be produced. Photons with the latter energies are evanescent and can therefore be neglected, which turns out to be equivalent to averaging the expression in time. We Fourier transform our current in time in order to use this discreteness and then, as we are only interested in effects in the probe, will later set the frequency $\omega$ to $\omega_{\trm{p}}$:
\bea 
\mbf{E}_{\trm{d}}(\mbf{r}_{\trm{d}}, \omega)\!&=&\!-\frac{1}{4 \pi} \int\! d^{3}r\;dt \Big[ \nabla \wedge 					\partial_{t}\mbf{M} -\nabla (\nabla \cdot \mbf{P}) 						+\partial_{t}^{2}\mbf{P} \Big] 
				\frac{\exp\big[-i\omega \big(|\mbf{r}_{\trm{d}} - \mbf{r}|+t\big)\big]}
                                    {|\mbf{r}_{\trm{d}} - \mbf{r}|}. \label{eqn:vacuum_cur_int}\qquad
\eea
It will be useful to expand the exponential using the assumption that the detector is placed much further away than the dimensions of the interaction volume, taken as the standard deviation width of the beams. Using $\trm{w}_{0,0} < \trm{w}_{\trm{p},0} \ll r_{\trm{d}}$, and then assuming $(\trm{w}_{0,0}/\lambda_{\trm{p}})(\trm{w}_{\trm{p},0}/r_{\trm{d}})^{2}, (\trm{w}_{\trm{p},0}/\lambda_{\trm{p}})(\trm{w}_{\trm{p},0}/r_{\trm{d}})^{3} \ll 1$, we can curtail the expansion to:
\bea \label{eqn:exp_approx}\gfunc{r_{\trm{d}}}{r} \approx \frac{1}{r_{\trm{d}}} 
                         \exp\Bigg\{-i\omega\Big(\big[r_{\trm{d}} - \mbf{\hat{r}_{\trm{d}}} \cdot \mbfr 
                                + \frac{1}{2r_{\trm{d}}} | \rhv \wedge \mbfr |^{2} 
                                 \big] + t \Big)\Bigg\}.
\eea
By retaining the quadratic co-ordinate terms, we indicate that we'll be working in the Fresnel regime. We can then split \eqnref{eqn:vacuum_cur_int} into three integrals and integrate by parts to remove surface terms, which we assume, using Gaussian expressions, tend to zero at the boundaries. This leaves us with:
\bea \label{eqn:Ed_simp}
\mbf{E}_{\trm{d}}(\mbf{r}_{\trm{d}}, \omega) \approx \frac{\omega^{2}\exp\big[-i\omega r_{\trm{d}}\big]}{4\pi r_{\trm{d}}}  \int & & \!\!\!\!\!d^{3}r\;dt\; \big( \mbf{M} \wedge \rhv + 					\mbf{P} -\mbf{P} \cdot \rhv \; \: \rhv \big)\times  \\
 & & \quad \exp\Big[i\omega \big( \mbf{\hat{r}_{\trm{d}}} \cdot \mbfr 
                                - \frac{1}{2r_{\trm{d}}} | \rhv \wedge \mbfr |^{2} -t\big)\Big].\nonumber
\eea
We expect the main vectorial contribution to the probe from the vacuum polarisation and magnetisation to be in the $x$ and $z$ directions, i.e. the directions of the probe and strong electromagnetic fields. 
When we substitute our particular scenario using \eqnref{eqn:pol_2} and \eqnref{eqn:mag_2} into the above equation and then Fourier transform back into $(\mbf{r}_{\trm{d}}, t)$ co-ordinates, we achieve the following:
\bea
\mbf{E}_{\trm{d}}(\mbfrd, t) & = & \mbf{E}^{\ast}_{\trm{d}}(\mbfrd)\frac{\exp[i(-\omega_{\trm{p}}r_{\trm{d}} + \omega_{\trm{p}}t + \psi_{\trm{p}})]}{2i} - \mbf{E}_{\trm{d}}(\mbfrd)\frac{\exp[-i(-\omega_{\trm{p}}r_{\trm{d}} + \omega_{\trm{p}}t + \psi_{\trm{p}})]}{2i},\nonumber\\
\mbf{E}_{\trm{d}}(\mbfrd) & := & \frac{I_{0}}{I_{\trm{cr}}} \frac{\alpha E_{\trm{p}}}{45 \lambda^{2}_{\trm{p}} r_{\trm{d}}} 
                           \left( (\mathcal{V}_{1} + \mathcal{V}_{2}) \mbf{u}_{1}
				     + (\mathcal{V}_{3}-\mathcal{V}_{4}) \mbf{u}_{2} 
				     + \Bigg(\sum_{i=1}^{4}\mathcal{V}_{i}\Bigg) \mbf{u}_{3} \right),\label{eqn:p12}
\eea
where the volumes $\mathcal{V}_{\trm{k}}$, and the vectors, $\mbf{u}_{i}$ are defined as the following:
\bea \label{eqn:Vk}
\mathcal{V}_{\trm{k}} := \!\int^{\infty}_{-\infty} \!\!&d^{3}r\:& \exp\Big[+i\omega_{\trm{p}} \Big(\frac{x^{2} + y^{2} + z^{2}}{2r_{\trm{d}}} 			- \frac{xx_{\trm{d}}+yy_{\trm{d}}+zz_{\trm{d}}}{r_{\trm{d}}} \\
& & \qquad-\frac{(xx_{\trm{d}} + yy_{\trm{d}} + zz_{\trm{d}})^{2} }{2 r^3_{\trm{d}}} +y\Big) - \frac{x^{2} + z^{2}}{\trm{w}_{\trm{p},0}^2} \Big]\nonumber \frac{\mathcal{I}_{\trm{k}}}{1+(z/z_r)^2} ;
\eea
\bea
\label{eqn:I_1} \mathcal{I}_{1} & := & \exp\Big[\!-\frac{2}{\trm{w}_0^2}\Big(x^{2} + y^{2} + a^{2} + b^{2}\Big)\Big] \exp\Big[\!-2i\Big(\omega_{0}z - \phi_{\trm{g},0}(z) + \frac{\omega_{0}z(x^{2}+y^{2} + a^{2} + b^{2})}{2(z^2 + z_r^2)}\Big)\Big], \nonumber \\
\label{eqn:I_2} \mathcal{I}_{2} & := & \exp\Big[\!-\!\frac{2}{\trm{w}_0^2}\Big(x^{2} + y^{2} + a^{2} + b^{2}\Big)\Big] \exp\Big[\!\phantom{-}2i\Big(\omega_{0}z - \phi_{\trm{g},0}(z) + \frac{\omega_{0}z(x^{2}+y^{2} + a^{2} + b^{2})}{2(z^2 + z_r^2)}\Big)\Big], \nonumber \\
\label{eqn:I_3} \mathcal{I}_{3} & := & \exp\Big[\!-\frac{2}{\trm{w}_0^2}\Big((x-a)^{2} + (y-b)^{2}\Big)\Big], \nonumber \\
\label{eqn:I_4} \mathcal{I}_{4} & := & \exp\Big[\!-\frac{2}{\trm{w}_0^2}\Big((x+a)^{2} + (y+b)^{2}\Big)\Big]; \nonumber
\eea
\bea
\mbf{u}_{1} & := & \left( \begin{array}{c}
                (1-\yrd)\cos\theta - \xrd(\xrd\cos\theta + \zrd\sin\theta) \\
                \zrd \sin\theta + \xrd\cos\theta - \yrd(\xrd\cos\theta + \zrd\sin\theta) \\
                (1-\yrd)\sin\theta - \zrd(\xrd\cos\theta + \zrd\sin\theta )
               \end{array} \right), \nonumber \\
\mbf{u}_{2} & := & \left( \begin{array}{c} 
                         \zrd\cos\theta + \frac{7}{4}\xrd\yrd\sin\theta\\
                        \frac{7}{4}( (\yrd)^2 - 1 )\sin\theta \\
                        -\xrd\cos\theta + \frac{7}{4}\yrd\zrd\sin\theta
                          \end{array} \right), \nonumber \\
\mbf{u}_{3} & := & \left( \begin{array}{c} 
                        (1-(\xrd)^2)\cos\theta \\
                        -\frac{7}{4}\zrd\sin\theta -\xrd\yrd\cos\theta \\
                        \frac{7}{4}\yrd\sin\theta -\xrd\zrd\cos\theta 
                          \end{array} \right). \nonumber
\eea
The main contribution from the integrals, $\mathcal{V}_{i}$, will be within the widths of our laser beams and so we can regard $\protect{x,y\lesssim\trm{w}_{0,0}},\; z\lesssim\trm{w}_{\trm{p},0}$. If we evaluate the expression at these values, the probe amplitude defocusing terms become $(1+(y/y_{\trm{r},\trm{p}})^2)^{-\frac{1}{2}} \approx 1-(1/2)(y/y_{\trm{r},\trm{p}})^{2}$, and we see that the correction $(1/2)(y/y_{\trm{r},\trm{p}})^{2} \ll \epsilon_{z}$ (the accuracy of our computation), and is therefore negligible. Moreover, considering the defocusing phase terms, when we assume these ranges for $x, y$ and $z$ throughout the integration, $\phi_{\trm{g},\trm{p}}(y) \approx (y/y_r) \ll 1$ and the final defocusing term $\omega_{\trm{p}}y(x^{2} + z^{2})/[2(y^{2}+y_{\trm{r},\trm{p}}^{2})] \approx 2\lambda_{\trm{p}}\trm{w}_{0,0}/(\pi \trm{w}_{\trm{p},0}^2)\ll1$ for the realistic parameters that we take for our lasers, defined later on \sxnref{sxn:analres}, \pref{ELIref}. Therefore, to be consistent with our beam expansion, we have considered all probe defocusing terms as constant ($\trm{w}_{\trm{p}}(y)\rightarrow\trm{w}_{\trm{p},0}$) within the integral above \eqnref{eqn:p12}. Whenever the probe occurs explicitly in expressions outside the integrals, the full space-dependence will be used.
\newline

From this integral \eqnref{eqn:Vk}, it can be seen that the $x$ and the $y$ co-ordinates can be integrated out to give just an integral in $z$ (see \eqnref{eqn:p12-in-1} in the appendix). On inspection, we notice certain factors in the complex exponential of the integrand constrain the diffracted field to be sharply peaked around $x_{\trm{d}}/r_{\trm{d}} \approx 0$ and $y_{\trm{d}}/r_{\trm{d}} \approx 1$, agreeing with physical intuition. 
Taking the limits $\mbfrd \rightarrow \mbf{y}_{\trm{d}}\equiv(0,y_{\trm{d}}, 0);\;a, b \rightarrow 0$, we can easily recover the expression for an on-axis measurement of a single strong beam + probe collision, given in \cite{dipiazza_PRL_06}.  Similarly, we can derive the diffracted magnetic field $\mbf{B}_{\trm{d}}(\mbf{r}_{\trm{d}},\omega_{\trm{p}})$, and using Maxwell's inhomogeneous equations again, show $\mbf{B}_{\trm{d}}(\mbf{r}_{\trm{d}},\omega_{\trm{p}}) = \mbf{\hat{k}}_{\trm{p}} \wedge \mbf{E}_{\trm{d}}(\mbf{r}_{\trm{d}},\omega_{\trm{p}})$ to within our calculational accuracy and i.e. that the flow of energy described by Poynting's vector goes as $\mbf{S}_{\trm{d}}(\mbf{r}_{\trm{d}},\omega_{\trm{p}}) = \mbf{E}_{\trm{d}}(\mbf{r}_{\trm{d}},\omega_{\trm{p}})  \wedge \mbf{B}_{\trm{d}}(\mbf{r}_{\trm{d}},\omega_{\trm{p}})/2 = |\mbf{E}_{\trm{d}}(\mbf{r}_{\trm{d}}, \omega_{\trm{p}})|^2\,\mbf{\hat{k}}_{\trm{p}} /2 $, which simplifies our calculation of the intensity pattern. We have assumed our earlier conditions on $\trm{w}_{0,0}$ and  $\trm{w}_{\trm{p},0}$ that we used in \eqnref{eqn:exp_approx}, as well as that $(y_{\trm{d}}/r_{\trm{d}})^{2}\approx1$.
\newline


One can question how sensitive these results are to being able to align the strong-field lasers parallel to one-another, by considering them being focused from a distance away by two large mirrors. For a small rotation $\delta\phi$ of $\mbf{k}_{0,1}$ and $\mbf{k}_{0,2}$ around the $x$-axis in the directions $\pm \mbf{\hat{y}}$ respectively, one can show for $\mbf{B}^{\prime}_{0,1}=\mbf{B}_{0,1}\cos\delta\phi + E_{0,1}\sin\delta\phi\,\mbf{\hat{z}}$, $\mbf{B}^{\prime}_{0,2}=\mbf{B}_{0,2}\cos\delta\phi - E_{0,2}\sin\delta\phi\,\mbf{\hat{z}}$ ($\mbf{E}^{\prime}_{0,1}=\mbf{E}_{0,1}, \mbf{E}^{\prime}_{0,2}=\mbf{E}_{0,2}$):
\bea
\mbf{P}(\delta\phi) & = & \mbf{P} - \delta\phi\frac{28\alpha^{2}}{45m^{4}}(E_{0,1}^{2}-E_{0,2}^{2})\cos\theta\,\mbf{\hat{y}} + O[(\delta\phi)^{2}], \\
\mbf{M}(\delta\phi) & = & \mbf{M} + \delta\phi\frac{16\alpha^{2}}{45m^{4}}(E_{0,1}^{2}-E_{0,2}^{2})\cos\theta\,\mbf{\hat{y}} + O[(\delta\phi)^{2}].
\eea
Keeping within the aforementioned bounds in the detector-plane, the corrections in $\delta\phi$ cancel in the combination $(\mbf{M} \wedge \rhv + \mbf{P} -\mbf{P}\!\cdot\!\rhv \; \rhv)$ meaning corrections to $\mbf{E}_{\trm{d}} \sim (\delta\phi)^2$, which implies, envisaging $\delta\phi\sim0.1$, that the parallel idealisation is sufficient to within the accuracy of the present treatment, $1/\pi$.
\newline

A further consideration would be what role the finite length of the strong-field beams plays during the passage of the probe beam. Both the diffracted intensity and polarisation effects that we will study are proportional to the intensity of the strong-field and therefore the corresponding longitudinal distribution is $\sim 1/(1+(z/z_{\trm{r},0})^{2})$. In the absence of a well-defined decay length, we take the effective length to be that at which the intensity falls to below a tenth of its initial value, giving an effective length of $l_{0}=3z_{\trm{r},0}=7.5\units{$\mu$m}$. A finite pulse length leads to consideration of the temporal envelope. For the case of a Gaussian beam, the leading temporal correction should be of the order $1/\omega_{0}\tau_{0}$ which we have already specified, through our assumption of monochromaticity, to be $\ll 1$. If we ensure that the strong-field pulse-length $\tau_{0}$ is such that $c\tau_{0} > 2l_{0}$ then the deviation should be negligible to within our level of accuracy. We therefore choose $c\tau_{0} \approx 2\times 2l_{0}$, with $\tau_{0}=100\units{fs}$, which will limit the maximum strong-field intensity obeying $I_{0}\tau_{0}A=\mathcal{E}$ for a fixed laser energy $\mathcal{E}$ and focus area $A$.

\section{Analysis of the results} \label{sxn:analres}
\setcounter{equation}{0}
\renewcommand{\theequation}{\Roman{section}.\arabic{equation}}
We present results that follow from the numerical evaluation in \matlab of the one-dimensional integral for $\mbf{E}_{\trm{d}}(\mbf{r}_{\trm{d}})$ given in \eqnref{eqn:p12-in-1} in the appendix. 
\newline

The results are presented in two sub-sections for i) intensity and ii) polarisation. These are further divided into the form of intensity along the x-axis, along the z-axis and in the x-z plane; following which we explain the polarisation rotation and ellipticity expressions along the probe propagation axis.

\subsection*{Intensity measurements off-axis}
\label{sxn:intens_oa}
\begin{figure}[!hp]
\noindent
\begin{center}
\includegraphics[width=0.5\linewidth]{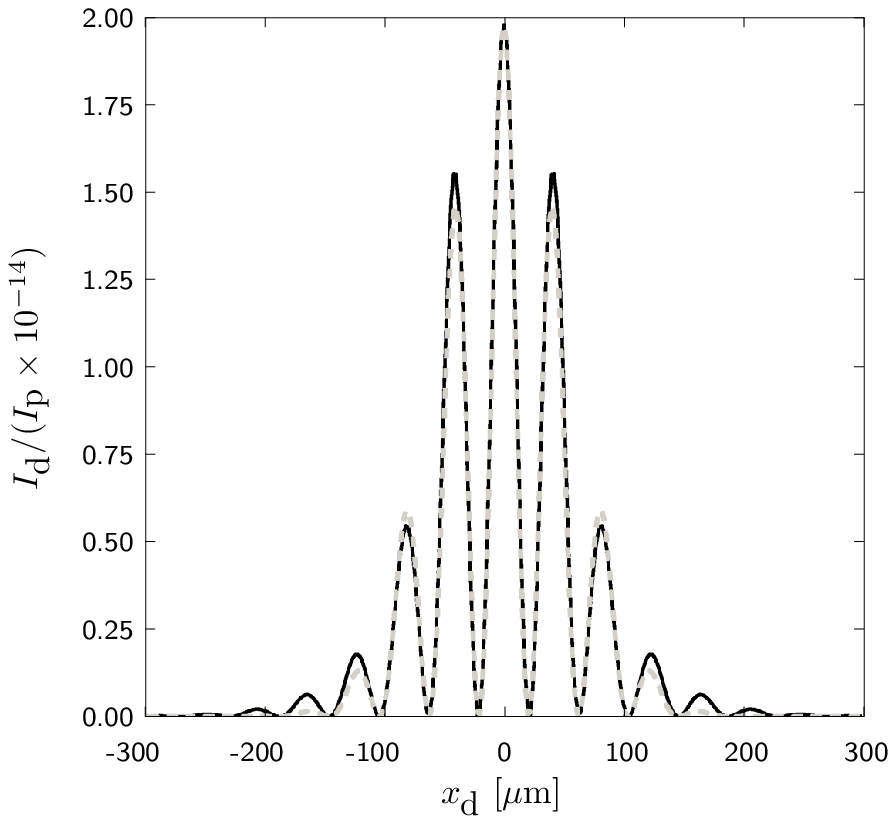}
\end{center}
\caption{The one-dimensional diffracted field along the $x$-axis that is predicted by first-order QED theory for the parameters: $a/\trm{w}_{0,0}=6, b/\trm{w}_{0,0}=0, \trm{w}_{0,0}=\lambda_{0}=0.8\units{$\mu$m}, \lambda_{\trm{p}}=0.4\units{nm}, \trm{w}_{\trm{p},0}=100\units{$\mu$m}, \theta=\pi/2, \trm{y}_{\trm{d}}=1\units{m}, I_{0} = 10^{24}\units{Wcm}^{-2}$, is plotted with a solid line. The dashed line indicates the result obtained by using the simplified analytical approach based on \eqnref{eqn:Id_approx}. \label{fig:phi_diff}} 
\begin{center}
\includegraphics[width=0.5\linewidth]{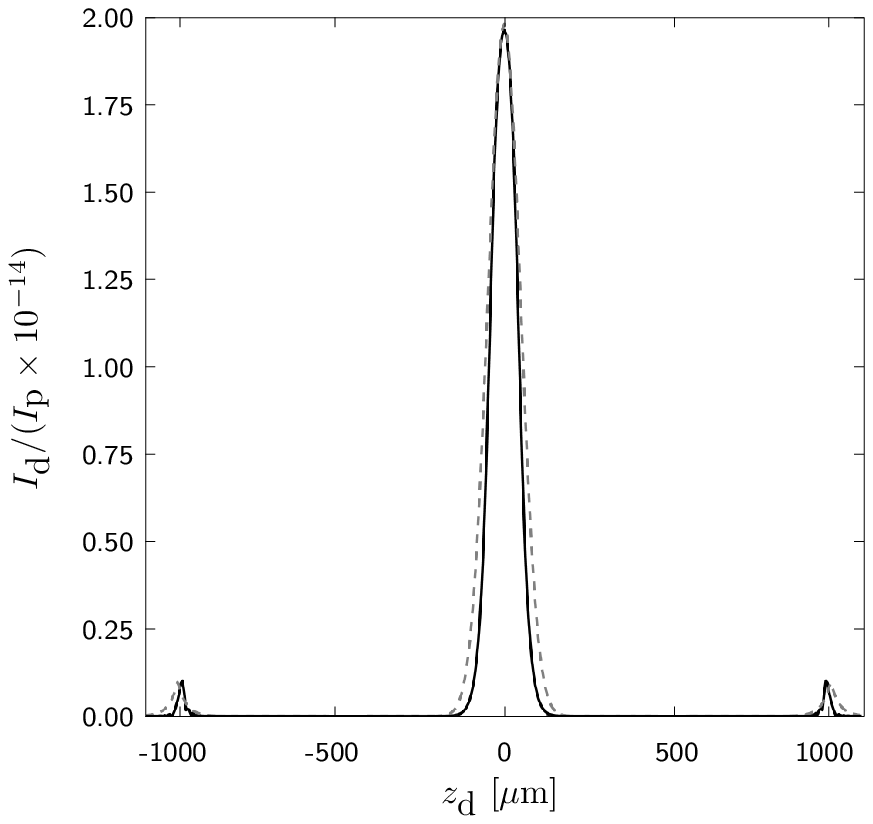}
\end{center}
\caption{The one-dimensional diffracted field along the z-axis i.e. along the axis of propagation of the strong-field beams, predicted by first-order QED theory is plotted with a solid line for the same physical parameters as in \figref{fig:phi_diff} but now with $a/\trm{w}_{0,0}=0$. The dotted line represents the simplified analytical approach keeping quadratic terms in the exponential, described in the text.\label{fig:theta_diff}}
\end{figure}
 The nonlinearity of the vacuum brought about by the two strong-field waves generates the diffraction patterns one would expect from a refractive solid-state material. The integral expression of $\mbf{E}_{\trm{d}}(\mbfrd, \omega)$ in \eqnref{eqn:Ed_simp} allows one to interpret the effect at hand as an example of Fresnel diffraction, including as it does, squared co-ordinate terms in the exponential. Satisfying the inequality: $(\trm{w}_{0,0}z_{\trm{r},0}/\lambda_{\trm{p}}r_{\trm{d}})(x^{2}_{\trm{d}}+z^{2}_{\trm{d}})/r_{\trm{d}}^{2} \ll 1$, we can neglect the $xz$ cross terms in the exponential and form two independent diffraction parameters, $\xi_{x} = \trm{w}_{0,0}^2/\lambda_{\trm{p}}r_{\trm{d}}$ and $\xi_{z} = \trm{w}_{\trm{p},0}^2/\lambda_{\trm{p}}r_{\trm{d}}$, for which $\xi_{x,z}\gg1$ implies taking the near-field limit, while $\xi_{x,z}\ll1$ implies we can take the far-field limit and hence the Fourier transform of the transmission function \cite{levi68}.

\subsubsection*{Analysis of field diffracted onto the $x$-axis}
Numerical evaluation of the leading-order QED contributions to the field diffracted onto the x-axis is shown in \figref{fig:phi_diff}. For our probe beam, we take the X-FEL at DESY in Hamburg, for which we have $80\units{GW}$ in a pulse of length $100\units{fs}$, of $0.4\units{nm}$ wavelength radiation focused into a waist $\trm{w}_{\trm{p},0}=100\units{$\mu$m}$ \cite{XFEL_TDR}. We maximise the intensity of the diffracted field by setting $\theta=\pi/2$. In addition, we take for our strong-field beams parameters from the upcoming ELI\label{ELIref} and HiPER facilities i.e. $\lambda_{0}=0.8\units{$\mu$m}$ and assume that they can be focused up to the diffraction limit i.e. that $\lambda_{0} = \trm{w}_{0,0}$ (although the consequences of focusing to a width of a few wavelengths are not drastic for the results). As already discussed, we then choose a pulse duration of $\tau_{0}=100\units{fs}$ to satisfy $c\tau_{0}\approx4 l_{0}$ and a total peak intensity of $I_{0}=10^{24}\units{Wcm}^{-2}$. The strong beams are separated by $a / \trm{w}_{0,0}=6$ and observations made at $y_{\trm{d}}=1\units{m}$. In \figref{fig:phi_diff}, we plot the diffracted field intensity, which clearly shows a familiar squared cosine, with a symmetric, decaying envelope function, similar to the square of the Fourier transform of a double-slit transmission aperture. This result is expected if one notes that with the above numerical parameters, the diffraction parameter $\xi_x$ along the $x$ direction is much smaller than unity. We also note at this point, that separation of the strong beams in the direction transverse to detector plate has in general no observable effect on our numerical results for intensity, which can be understood intuitively. As the vacuum signal $\mbf{E}_{\trm{d}}$ is created in phase with the probe, the total phase difference between sources of vacuum waves separated by $2b$ in the longitudinal direction is $2b\omega_{\trm{p}}(1-y_{\trm{d}}/r_{\trm{d}})$, as can be seen from \eqnref{eqn:exp_approx} in the far-field. Setting $z_{\trm{d}}=0$ for simplicity, the condition to be fulfilled for a corresponding first minimum would be $\lambda_{\trm{p}}/2 = b(x_{\trm{d}}/r_{\trm{d}})^{2}$. Since we are using an X-ray probe, for realistic separation of the strong-field beams of the order of a few multiples of $\trm{w}_{0,0}$, the first minimum would occur for values of $x_{\trm{d}}$ far outside our detector plate. Moreover, for a finite strong-beam $x$-separation $2a$, any additional separation of the beams in the $y$-direction, will neither create an appreciable separation perpendicular to the diffracted wave vector. For these reasons, we can disregard $b$ and set it equal to zero in this section.
\newline 

In the present case $x_{\trm{d}},z_{\trm{d}}\ll y_{\trm{d}}$ and this implies that the terms proportional to the vectors $\mathbf{u}_1$ and $\mathbf{u}_2$ in $\mbf{E}_{\trm{d}}(\mbf{r}_{\trm{d}})$ are negligible. Also, we notice that for the typical situation, $\trm{w}_{0,0}+a \ll \trm{w}_{\trm{p},0}$, the cosine term formed from the integrals $\mathcal{V}_{1} + \mathcal{V}_{2}$ can be neglected when both:
\bea
\frac{2\pi\trm{w}_{\trm{p},0}}{\lambda_{0}} \gg \sqrt[4]{1 + \Bigg(\frac{\pi \trm{w}_{\trm{p},0}^{2}}{ y_{\trm{d}}\lambda_{\trm{p}}}\Bigg)^{2}}\qquad\trm{and}\qquad \frac{\lambda_{0}}{2\lambda_{\trm{p}}}\frac{z_{\trm{d}}}{r_{\trm{d}}}\ll 1.\label{eqn:V1cond}
\eea
These observations then considerably reduce our diffraction integral in \eqnref{eqn:p12} to just:
\crampest
\bea \label{eqn:p12approx}
\mbf{E}_{\trm{d}}(\mbf{r}_{\trm{d}})&\!\approx&\!\frac{\alpha}{45 \lambda_{\trm{p}}^{2} r_{\trm{d}}}                          \frac{I_0}{I_{\trm{cr}}}E_{\trm{p}}(\mathcal{V}_{3} + \mathcal{V}_{4})\mbf{u}_{3}.
\eea \uncramp

The full Fresnel-like diffraction integral which couples the $x$, $y$ and $z$ co-ordinates together, is unwieldy when attempting to garner qualitative information. Assuming $\xi_{x} \ll 1$, the Fresnel integral will produce a diffraction pattern with the same shape as if we took the Fourier limit. In this way, by performing the integral in $x$ in $\mathcal{V}_{3}$ and $\mathcal{V}_{4}$ we obtain for the diffracted field intensity, $I_{\trm{d}}(\mathbf{r}_{\trm{d}},\omega_{\trm{p}})=|\mbf{S}_{\trm{d}}(\mbf{r}_{\trm{d}},\omega_{\trm{p}})|$ that:
\bea
I_{\trm{d}}(x_{\trm{d}}, y_{\trm{d}}, z_{\trm{d}}=0,\omega_{\trm{p}})\sim I_{\trm{p},0}\exp\Big[-\frac{(x_{\trm{d}}/r_{\trm{d}})^{2}}{2\sigma_{\trm{d},x}^{2}}\Big]\cos^{2} \Big[\omega_{\trm{p}}a(x_{\trm{d}}/r_{\trm{d}})\Big]; \quad
\sigma^{2}_{\trm{d},x} := \frac{\lambda_{\trm{p}} \sqrt{2}}{\pi \trm{w}_{0,0}},\label{eqn:Id_approx}
\eea
with $I_{\trm{p},0}=E_{\trm{p}}^2/2$, which is what one would expect from the Fourier transform of a Gaussian convoluted with two delta functions. The cosine term originates from the interference between the vacuum current generated in the two slits, and the Gaussian is the effect of the single-slit shapes of both the strong beams. We use the $\sim$ sign to emphasise the illustrative nature of our arguments, as although the fringe positions are correctly predicted, the single-slit shape is incorrect, as seen in \figref{fig:phi_diff}. This is an example of a consequence of non-trivial beam geometry, for which the full three-dimensional integration must be performed.

\subsubsection*{Analysis of field diffracted onto the $z$-axis} \label{sxn:z-diffraction}
An example of a diffraction pattern in the $z$-direction is shown in \figref{fig:theta_diff}. The numerical parameters are those used in the above case but with $a / \trm{w}_{0,0}=0$ and now the reverse situation $x_{\trm{d}}=0$ and $z_{\trm{d}}\ll y_{\trm{d}}$. From this figure, we see that the intensity pattern is formed by a central peak, of width $\approx 50\units{$\mu$m}$, and two smaller exponential-shaped peaks some distance away. Concerning the central peak, when we consider that the amplitude of the strong field along the $z$-axis, and hence the ``vacuum transmission aperture'' is governed by the factor $1/(1+(z/z_{\trm{r},0})^{2})$, we see very clearly that the diffracted electromagnetic field does not result from the aperture's Fourier transform, which would have been a decaying exponential, symmetric about the origin, i.e. the wrong shape and with a smaller width of about $10\units{$\mu$m}$. The presence of the two peaks can be described by the diffraction-grating-like sinusoid along the $z$-axis. That the simple Fourier analysis applied in the previous case does not work here, is already clear from the diffraction parameter $\xi_z\sim25$ not being smaller than unity.
\newline

We wish to again explain our diffracted field qualitatively, but now how it, and so how \eqnref{eqn:p12}, depends upon the $z$ co-ordinate in the detector plane, $z_{\trm{d}}$. We can see from  \eqnref{eqn:p12-in-1} how the decay of the integrand in the $z$-direction is controlled by the softcore $1/(1+(z/z_{\trm{r},0})^{2})$ term. The importance and presence of this term prevents us separating out the $z$ from the $x$ and $y$ integration variables and hence the $z_{\trm{d}}$ from the $x_{\trm{d}}$ and $y_{\trm{d}}$ detector-plane co-ordinates in the integrands $\mathcal{V}_{\trm{k}}$, as we managed to do in the previous case. However, if we consider that $x_{\trm{d}}=0$ and that $z_{\trm{d}}\ll y_{\trm{d}}$, we can again neglect in Eq. (\ref{eqn:Ed_simp}) the terms proportional to the vectors $\mathbf{u}_1$ and $\mathbf{u}_2$. Unlike the previous situation however, the condition \eqnref{eqn:V1cond} to neglect the integrals $\mathcal{V}_{1}$ and $\mathcal{V}_{2}$ is not fulfilled for arbitrary $z_{\trm{d}}\ll r_{\trm{d}}$ and they are accordingly not negligible. It can be shown by performing an analysis similar to the one in the previous case that the integrals $\mathcal{V}_{3}$ and $\mathcal{V}_{4}$ give rise to the central peak (with width $\trm{w}_{\trm{p},0}/2=50\units{$\mu$m}$) while the integrals $\mathcal{V}_{1}$ and $\mathcal{V}_{2}$ give rise to the secondary smaller peaks located at $z_{\trm{d}} = \mp2 r_{\trm{d}}\lambda_{\trm{p}}/\lambda_{0}=\mp 1000\units{$\mu$m}$. Therefore, the secondary peaks originate from the standing wave of the strong field, which the probe experiences as if it were a diffraction grating, and is another example of the effect of non-trivial beam shape. Similar arguments leading to \eqnref{eqn:Id_approx}, retaining the quadratic terms in the exponential give the dashed line in \figref{fig:theta_diff} and again show good agreement.

\subsubsection*{Single-slit pattern}
We have now seen from some results that a consequence of the non-trivial strong-field beam shape is a deviation from the ideal double-slit analogy. As mentioned in the introduction, this mainly affects the interpretation of each strong-field laser as a single slit. We can illustrate the difference brought about by smooth edges when we consider diffraction from a single slit of dimension $2l_{x}$ by $2l_{z}$, centred at the origin. The diffracted electric field in the far-field, $E_{\trm{d},\trm{Rect}}$ can be calculated via Fourier transformation of the aperture function:
\begin{equation}
E_{\trm{d},\trm{Rect}}\propto \int_{-\infty}^\infty dx \int_{-\infty}^\infty dz \exp\left(-i\omega\xrd x - i\omega\zrd z\right)\trm{Rect}\left(\frac{x}{l_{x}}\right)\trm{Rect}\left(\frac{z}{l_{z}}\right),
\end{equation}
where $\trm{Rect}(x/a)$ equals unity only in the region $x\in\,]-a,a[$, being otherwise zero. This gives a diffracted intensity $I_{\trm{d},\trm{Rect}}\propto|E_{\trm{d},\trm{Rect}}|^{2}$:
\begin{equation}
I_{\trm{d},\trm{Rect}}\propto \frac{\sin^{2} (\omega x_{\trm{d}}/r_{\trm{d}})}{(\omega x_{\trm{d}}/r_{\trm{d}})^{2}}\frac{\sin^{2} (\omega z_{\trm{d}}/r_{\trm{d}})}{(\omega z_{\trm{d}}/r_{\trm{d}})^{2}},
\end{equation}
which gives the familiar single-slit minimum conditions $(n+1/2)\lambda=2l_{x}x_{\trm{d},n}/r_{\trm{d}}$, $(n+1/2)\lambda=2l_{z}z_{\trm{d},n}/r_{\trm{d}}$, for $n\in\mathbb{Z}$ and $\lambda=2\pi/\omega$. For our ``Gaussian'' slits, our diffracted electric field $E_{\trm{d},\trm{Gauss}}$ becomes:
\begin{equation}
E_{\trm{d},\trm{Gauss}}\propto \int_{-\infty}^\infty dx \int_{-\infty}^\infty dz \exp\left(-i\omega\xrd x- i\omega\zrd z\right)\exp\left(-\frac{x^{2}}{\trm{w}_{0,0}^{2}(1+z^{2})}\right)\frac{1}{1+z^{2}}.
\end{equation}
This can be analytically evaluated after setting $z_{\trm{d}}=0$, giving an intensity:
\begin{equation}
I_{\trm{d},\trm{Gauss}}\propto \exp\Bigg[-\Big(\frac{\omega \trm{w}_{0,0}x_{\trm{d}}}{2r_{\trm{d}}}\Big)^{2}\Bigg]K^{\,2}_{0}\Bigg[\frac{1}{2}\Big(\frac{\omega \trm{w}_{0,0} x_{\trm{d}}}{2r_{\trm{d}}}\Big)^{2}\Bigg],
\end{equation}
where $K_{0}$ is the zeroth-order modified Bessel function of the first kind and is monotonically decreasing, i.e. without fringe structure. As other terms introduce only a finer structure and as the final integration in $y$ would also be over a smooth function, we see that no periodicity arises from our single-slit diffraction pattern, which is consistent with numerical results. Beyond the far-field limit however, an interference-like deviation would be expected to develop. One example of this was calculated in \cite{mocken03}, where a relativistic Gaussian electron wave-packet in the Coulomb field of some highly-charged ions acquires an interference pattern structure when placed in an intense laser field.

\subsubsection*{Resultant intensity difference off-axis} \label{sxn:intensity_offaxis}
For the relevance to experiment however, instead of just plain diffraction theory, we will be more interested in studying the \emph{difference} brought about by vacuum polarisation effects. With $\langle\rangle$ denoting an average over a laser cycle, the difference can be shown to be:
\bea
I_{\trm{tot}} - I_{\trm{p}} &=&\!\!\Big(\langle|\mbf{E}_{\trm{p}} +\mbf{E}_{\trm{d}}|^{2}\rangle - \langle|\mbf{E}_{\trm{p}}|^{2}\rangle \Big) \nonumber\\
&=&\!\! I_{\trm{pd}}+I_{\trm{d}}, \nonumber\\[1ex]
I_{\trm{pd}} &=&\!\!\frac{I_{0}}{I_{\trm{cr}}} \frac{\alpha I_{\trm{p},0}}{180\pi \lambda_{\trm{p}}^{2}} \frac{ \exp\!\Big[\!\!-\!\!\big(x_{\trm{d}}^{2} + z_{\trm{d}}^{2}\big)/\trm{w}_{\trm{p}}^2\Big]}{y_{\trm{d}} \sqrt{1+(y_{\trm{d}}/y_{\trm{r}})^2}} \big(\bm{\mathcal{V}}^{i}\sin\eta\!-\!\bm{\mathcal{V}}^{r}\cos\eta\big)\!\cdot\!\left( \mbf{\hat{x}}\cos\theta\!+\!\mbf{\hat{z}}\sin\theta \right),\qquad\\[0.5ex]
I_{\trm{d}} &=&\langle| E_{\trm{d}} |^{2} \rangle;
\eea
\bea \label{eqn:intensity_total_simple}
&\quad& \bm{\mathcal{V}}^{r} := \trm{real}(\bm{\mathcal{V}}), \qquad
\bm{\mathcal{V}}^{i} := \trm{imag}(\bm{\mathcal{V}}), \\
\bm{\mathcal{V}} & = & (\mathcal{V}_{1} + \mathcal{V}_{2})\mbf{u_{1}} 
		     +  (\mathcal{V}_{3} - \mathcal{V}_{4})\mbf{u_{2}}
		     +  \left(\sum_{i=1}^{4} \mathcal{V}_{i}\right) \mbf{u_{3}}, \\
&\quad& \eta = \tan^{-1}\left(\frac{y_{\trm{d}}}{y_{\trm{r}}}\right) - \frac{\omega_{\trm{p}}y_{\trm{d}}}{2}\frac{x_{\trm{d}}^{2} + z_{\trm{d}}^2}{y_{\trm{r}}^2 + y_{\trm{d}}^2}.
\eea

\begin{figure}[!hp]
\noindent
\begin{center}
\includegraphics[width=0.5\linewidth]{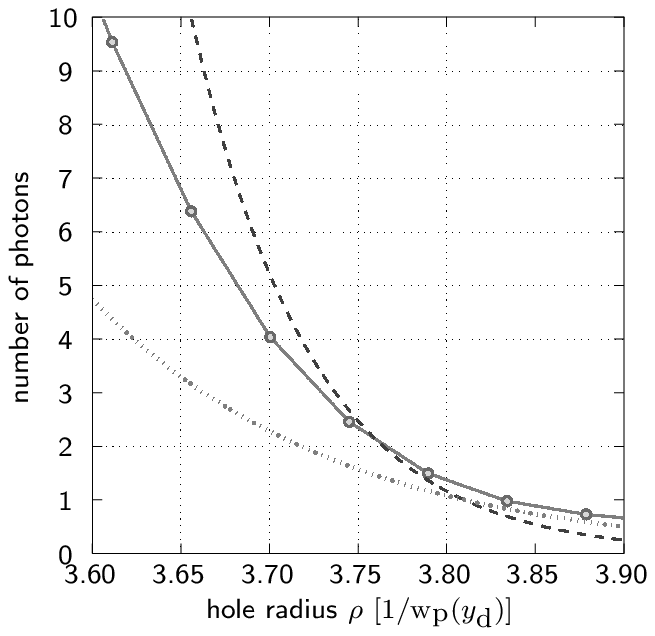}\\[3ex]
\end{center}
\caption{For a probe beam focused to $\trm{w}_{\trm{p},0}=3.6~\mu\trm{m}$, $\trm{w}_{\trm{p}}(y_{\trm{d}})=5.0~\mu\trm{m}$ and a large enough detector hole radius $\rho$, the photon count from the vacuum-probe cross term $N_{\trm{pd}}(\rho)$ (solid-line) becomes comparable to that from the probe $N_{\trm{p}}(\rho)=N_{\trm{p}}(0)\exp(-2(\rho/\trm{w}_{\trm{p}}(y_{\trm{d}}))^{2})$ (dashed-line) (here around $2$ diffracted photons) and greater than statistical noise from the probe $\sqrt{N_{\trm{p}}(\rho)}$ (dotted-line).}
\label{fig:ph_count}
\end{figure}

We can evaluate this expression on an $x_{\trm{d}}$--$z_{\trm{d}}$ grid at a fixed distance $y_{\trm{d}}$, and calculate the differences in photon rates brought about by the polarised vacuum.
\begin{figure}[!hp]
\noindent
\begin{center}
\includegraphics[width=0.7\linewidth]{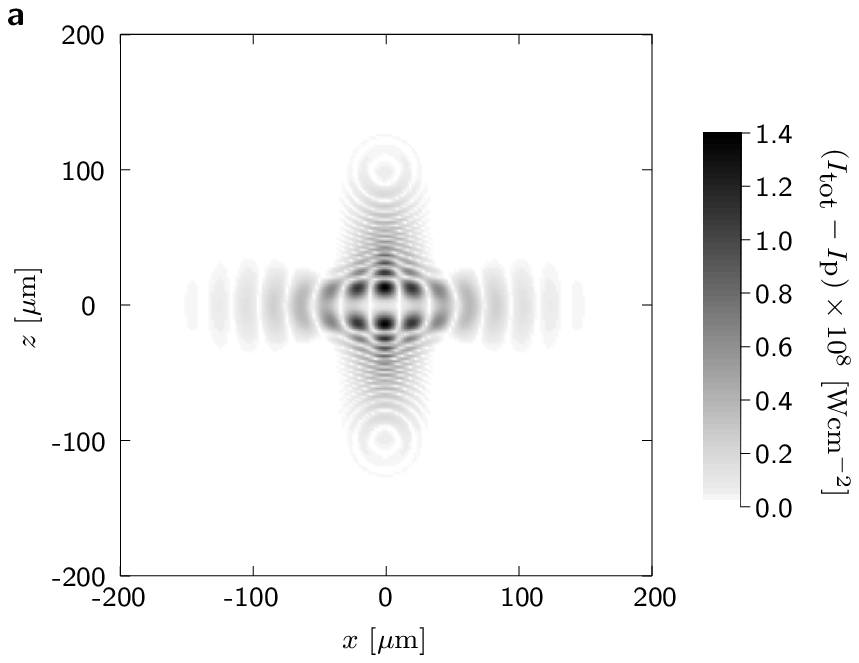}\\
\includegraphics[width=0.7\linewidth]{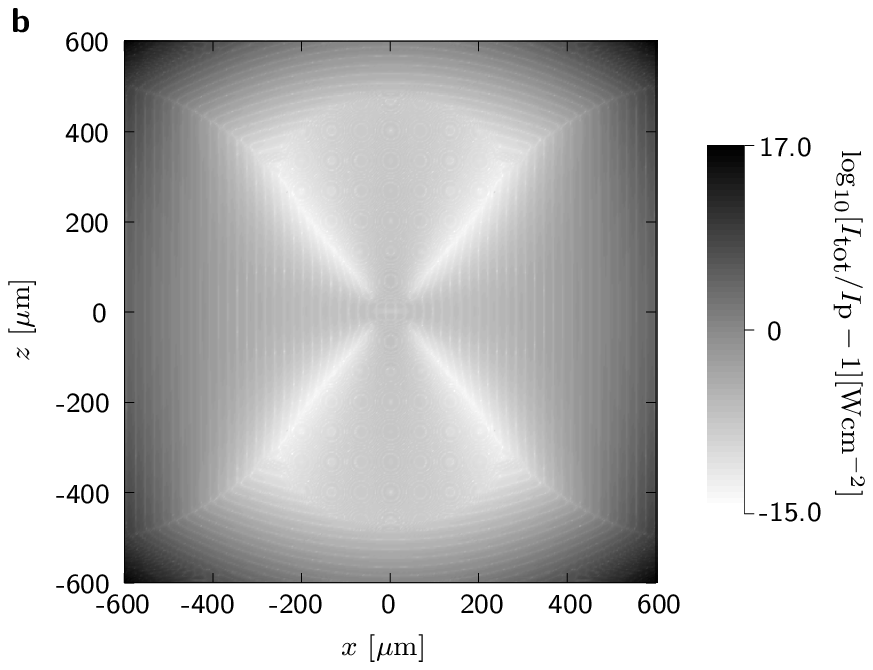}
\end{center}
\caption{Plot \textsf{\textbf{a}} is of the quantity $I_{\trm{pd}}$, the vacuum-probe cross-term in time-averaged total intensity, at a distance $r_{\trm{d}}\approx y_{\trm{d}} = 1\units{m}$, with the same experimental parameters as in the example of diffraction along the x-axis apart from $a/\trm{w}_{0,0}=12$. Plot \textsf{\textbf{b}} is the logarithm to base 10 of this divided by the time-averaged probe intensity i.e. the logarithm of the signal-to-noise ratio for the same parameters.}\label{fig:grid_intensity}
\end{figure}
Our procedure was to make many such grids that became ever-finer, so that we could see how the integral and i.e. the predicted number of photons per shot converged. Since the diffraction pattern must be smooth, this number should then yield a reliable value. We took the same following parameters of a typical experimental run: $y_{\trm{d}}=1\units{m}, a/\trm{w}_{0,0}=12, b/\trm{w}_{0,0} = 0, \trm{w}_{0,0} = \lambda_{0} = 0.8\units{$\mu$m}, I_{0}=10^{24}\units{Wcm}^{-2}$, giving the patterns shown in \figrefa{fig:grid_intensity} (the parameters of the probe field were those already employed in the previous examples). We focus on the diffracted photons described by the interference term between $\mathbf{E}_{\trm{p}}(\mathbf{r}_{\trm{d}},t)$ and $\mathbf{E}_{\trm{d}}(\mathbf{r}_{\trm{d}},t)$. This term spreads out in the $x\text{-}z$ plane with a width $\sqrt{2}$ larger than that for the probe field, as the multiplying diffraction signal has a much wider overall decay, so there exist regions in which the ratio of diffracted to probe signal is favourable which can be seen on the log-plot of the total difference due to vacuum signal over the probe background \figrefb{fig:grid_intensity}. At the same time, moving too far from the centre of the pattern will reduce the intensity to the point where nothing can be detected. If we consider drilling a hole of radius $\rho$ into the centre of the detector and approximate the decay of $I_{\trm{pd}}$ to come entirely from the probe Gaussian, considering the single strong-beam scenario in order to maximise signal, we can obtain limits on $\rho$:
\bea
\Big[\ln\frac{N_{\trm{p}}}{N_{\trm{pd}}} \Big]^{1/2}  \lesssim \frac{\rho}{\trm{w}_{\trm{p}}(y)} \lesssim \Big[\ln N_{\trm{pd}} \Big]^{1/2},
\label{eqn:rho}
\eea
for total incident probe and cross-term diffracted photons $N_{\trm{p}}$, $N_{\trm{pd}}$. This agrees with the intuitive notion that to stand any chance of measurement, the signal must be larger than statistical noise from the background, which if modelled with Poisson statistics implies $N_{\trm{pd}}>\sqrt{N_{\trm{p}}}$ \footnote{When the statistical error on the number of photons is modelled by a Poisson distribution, the relative error in the mean photons measured per shot, $\mu$, is given by $1/\sqrt{n\mu}$ for $n$ shots. As long as lower intensity lasers still satisfy the condition $N_{\trm{pd}}>\sqrt{N_{\trm{p}}}$, they can indeed be used, it is just a question of how long the experiment can be run to make $n\mu$ large enough to be certain to have observed an effect.}. We can either fulfil this condition that the vacuum signal is larger than the minimum background noise over the entire plate, or we can consider measuring counts only in regions where $N_{\trm{pd}}(\rho)\gtrsim N_{\trm{p}}(\rho)$. In both cases, the number of diffracted photons will simply increase with probe intensity, whereas as $N_{\trm{p}}$ depends only on the probe laser energy and wavelength $\lambda_{\trm{p}}$ and so for larger probe intensity, we can easier fulfil both bounds on $\rho$ in \eqnref{eqn:rho}.  First setting $y_{\trm{d}}=50\units{cm}$ and the still at ELI comfortably attainable $I_{0}=5\times 10^{24}\units{Wcm$^{-2}$}$, for a probe focal width of $8~\mu\trm{m}$, we achieve $N_{\trm{pd}}=7.5\times10^{7}$ from $N_{\trm{p}}=8.0\times10^{12}$ probe photons per shot. Secondly, we can plot how $N_{\trm{pd}}(\rho)$ varies with hole radius, and for a tighter probe beam focal width of $\trm{w}_{\trm{p},0}=3.6~\mu\trm{m}$, which, in the light of recent results of focusing hard $20~\trm{keV}$ photons to a width of $7~\trm{nm}$ \cite{mimura09}, we expect to be attainable in the near future, we achieve the dependency shown in \figref{fig:ph_count}. In the region $N_{\trm{pd}}(\rho)\gtrsim N_{\trm{p}}(\rho)$, taking into account the efficiency of commercially-available CCDs for $\lambda_{\trm{p}}=0.4\units{nm}$ or $3.1\units{keV}$ photons ($\gtrsim90\%$ \cite{CCD_site}), we expect approximately two diffracted photons to be measurable per shot of the probe beam. This can then be compared with results for $I_{\trm{d}}$ which, by not being subject to the probe Gaussian envelope, has a much wider spread, and is possibly easier to measure as reported in \cite{king10a}, with the caveat that an optical probe beam was used with a total energy $2.5\times10^{3}$ larger than in the present X-ray case.

\subsection*{Polarisation results (double-slit)}
This section concerns itself with the induced ellipticity and rotation of the probe polarisation due to VPEs, that can be measured on the probe beam's propagation axis. Setting $x_{\trm{d}}=z_{\trm{d}}=0$,
it can be shown that new expressions that incorporate defocusing terms in the probe, for the polarisation, $\psi$ and ellipticity, $\eps$, are given by:
\protect\bea
\label{eqn:psi} \psi & = & \frac{\alpha \sin2\theta}{120\lambda_{\trm{p}}^{2}} \frac{I_0}{I_{\trm{cr}}} 
	   \sum_{k=1}^{4} \left(\frac{\mathcal{V}^i_{\trm{k}}}{y_{\trm{r},\trm{p}}} + \frac{\mathcal{V}^r_{\trm{k}}}{y_{\trm{d}}}  \right),\\
\label{eqn:eps} \eps & = & \frac{\alpha \sin2\theta}{120\lambda_{\trm{p}}^{2}} \frac{I_0}{I_{\trm{cr}}} 
	   \sum_{k=1}^{4} \left( \frac{\mathcal{V}^r_{\trm{k}}}{y_{\trm{r},\trm{p}}} - \frac{\mathcal{V}^i_{\trm{k}}}{y_{\trm{d}}}\right),
\eea
where in the limit of $a, b, \rightarrow 0; \: y_{\trm{r},\trm{p}} \rightarrow \infty$, we again recover the expression in the original paper \cite{dipiazza_PRL_06}. We also note that the introduction of experimentally relevant defocusing terms in the probe, produces the more realistic and expected result that  $\lim_{y_{\trm{d}}\rightarrow\infty}\{\psi,\eps\}\neq\{0,0\}$.
\newline

\begin{figure}[!hp]
\noindent
\begin{center}
\includegraphics[width=0.5\linewidth]{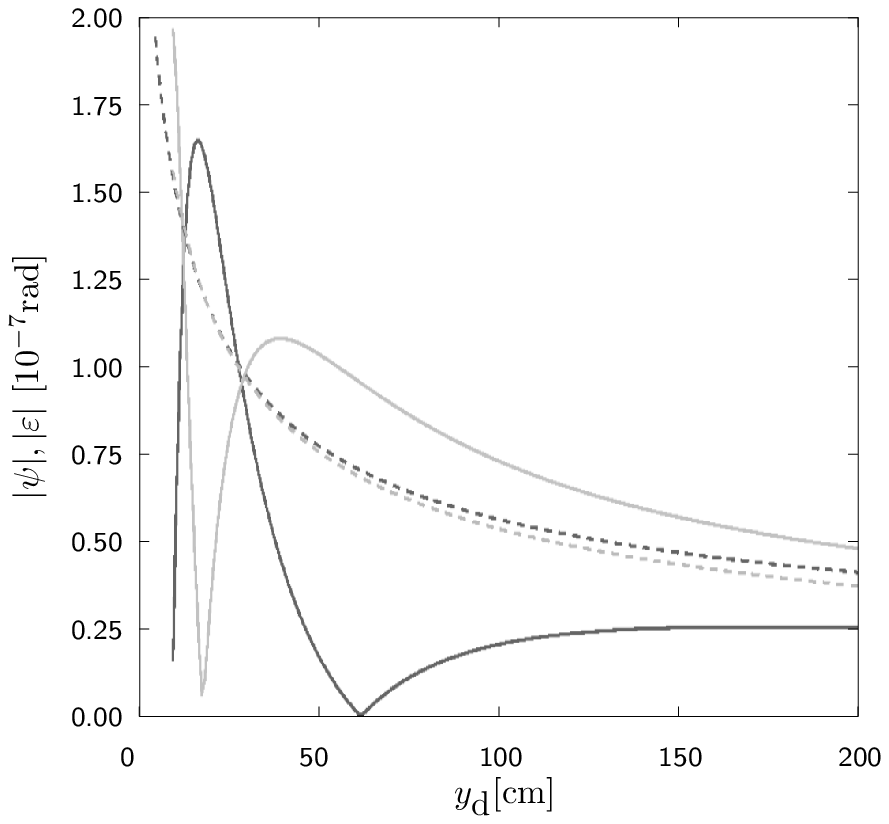}\\[3ex]
\end{center}
\caption{The absolute value of the new polarisation and ellipticity expressions derived with probe defocusing terms and separated strong field beams as a function of the observation distance $y_{\trm{d}}$, for the same parameters as in Fig. (\ref{fig:phi_diff}) but with $a/\trm{w}_{0,0}=12$. The dashed lines are the former analytical polarisation and ellipticity formulae for $a=0$ without probe defocusing terms, given in \cite{dipiazza_PRL_06}, with the solid lines the new polarisation and ellipticity presented in this paper. Darker lines are drawn for polarisation rotation and lighter ones for ellipticity\label{fig:psi_v_rd}} 
\end{figure}
Varying $y_{\trm{d}}$ with $x_{\trm{d}}/r_{\trm{d}}, z_{\trm{d}}/r_{\trm{d}} = 0$, and setting $\theta=\pi/4$ to maximise the effect of the polarised vacuum, we show a demonstrative plot in \figrefa{fig:psi_v_rd}, for how $\psi$ and $\eps$ vary for a fixed strong-field beam separation, $a/\trm{w}_{0,0}=12$, with the other parameters the same as in the previous examples. The first difference we note is that in comparison with results from \cite{dipiazza_PRL_06}, for $y_{\trm{d}} \ll y_{\trm{r},\trm{p}}$, polarisation and ellipticity oscillate rapidly and there are sizeable ranges where both are larger than that for previously derived results. For the choice of parameters in the plot, $y_{\trm{r},\trm{p}}\approx80\units{m}$, and so if we keep within this range, i.e. disregard the effect of defocusing terms, we can clearly ascertain the improvement brought by separating the strong-field beams. This perhaps counterintuitive result can be shown to be consistent with our analysis by following through these conditions on the detector-plane co-ordinates and studying the form of the integrals $\mathcal{V}_{3,4}$ which appear in our expressions for $\psi$ and $\eps$ (we can once more disregard the contribution of $\mathcal{V}_{1,2}$):
\bea 
\mathcal{V}_{3,4}\!\!&=&\!\! \int_{-\infty}^{\infty}\!\! dz~\frac{1}{1+(z/z_{\trm{r},0})^{2}}\exp\!\Big[\!-\!i\omega_{\trm{p}}\frac{z^{2}}{2y_{\trm{d}}}\!-\!\frac{z^{2}}{\trm{w}^{2}_{\trm{p},0}}\Big]\mathcal{I}_{y,\pm}\, \mathcal{J}_{x,\pm}, \label{eqn:eps_psi_approx}\\
\mathcal{I}_{y,\pm}\!\!&=&\!\!\int_{-\infty}^{\infty}\!\! dy~ \exp\!\Big[\frac{-2(y-b)^{2}}{\trm{w}^{2}_{0}}\Big],\label{eqn:eps_psi_approx_I_part}\\
\mathcal{J}_{x,\pm}\!\!&=&\!\!\int_{-\infty}^{\infty}\!\! dx~ \exp\!\Big[\!-\!i\omega_{\trm{p}}\frac{x^{2}}{2y_{\trm{d}}} \Big] \exp\!\Big[\!-\!\frac{x^{2}}{\trm{w}^{2}_{\trm{p},0}}\Big]\exp\!\Big[ \frac{-2\big(x\mp a\big)^{2}}{\trm{w}^{2}_{0}}  \Big]\label{eqn:eps_psi_approx_J_part}.
\eea
From \eqnref{eqn:eps_psi_approx_I_part} we can see more clearly, that under these conditions (especially as $y_{\trm{r},\trm{p}}\gg 2(\trm{w}_{0,0}+b)$), since there is no other structure in the $y$-direction, $b$ becomes an inconsequential parameter when measuring polarisation and ellipticity, just as it was for the diffracted field, and will likewise be set to zero. By separating strong-field beams in the $x$-direction, we see that we only produce an effect on the $x$-integrals, $\mathcal{J}_{x,\pm}$. When considering the contribution from the first complex exponential factor in \eqnref{eqn:eps_psi_approx_J_part}, for a fixed $a$, in varying $y_{\trm{d}}$, we vary the overlap this factor's real cosine and imaginary sine functions with the other two Gaussian integrand factors, which have maxima at $x=0$ and $x=\mp a$ respectively. Hence some values of $y_{\trm{d}}$ form local maxima in $\psi$ and $\eps$, and due to the trigonometric nature of the varying function, we have the oscillating shape in \figref{fig:psi_v_rd}. However, in the limit $y_{\trm{d}}\rightarrow0$, (taking into account all $\mathcal{V}$'s), both of these values tend to constants:
\protect\bea
\psi = 0; \quad
\eps = \frac{\alpha\sqrt{\pi}}{30\sqrt{2}}\frac{I_{0}}{I_{\trm{cr}}}\frac{\trm{w}_{0,0}}{\lambda_{\trm{p}}}\exp\Big(\frac{-2a^{2}}{\trm{w}_{0,0}^{2}}\Big)\sin2\theta.
\eea
\begin{figure}[!hp]
\noindent
\begin{center}
\includegraphics[width=0.5\linewidth]{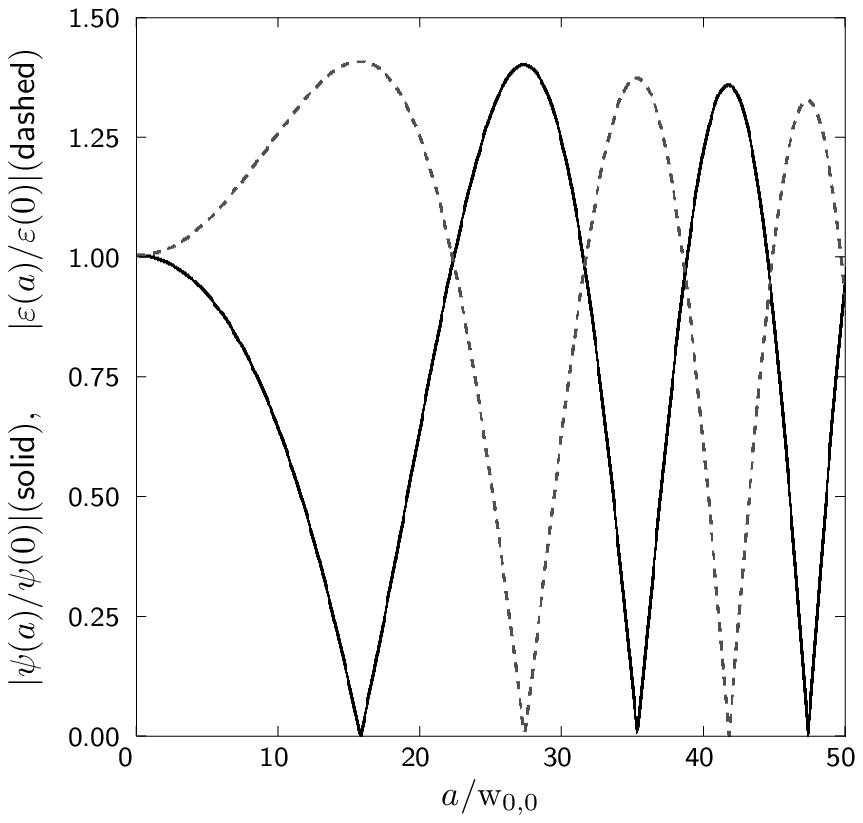}
\end{center}
\caption{The ratios of $|\psi(a) / \psi(0)|$ (continuous line) and $|\eps(a) / \eps(0)|$ (dashed line) in the double-slit set-up for the same parameters as in Fig. (\ref{fig:phi_diff}) but with $\trm{w}_{\trm{p},0}=200\units{$\mu$m}$. The ratio $|\psi(a)/\psi(0)|$ was chosen in preference to the ratio of $|\psi(a)|$ with that in the original paper \cite{dipiazza_PRL_06} as introduction of focusing terms doesn't make the latter a viable comparison. \label{fig:psi_v_a}}
\end{figure}
\noindent We also show how $\psi$ and $\eps$ depend upon beam-separation $a$ in \figref{fig:psi_v_a}, and can show consistency by using the same arguments as above for the dependence on $y_{\trm{d}}$. In varying $a$, the first two factors in \eqnref{eqn:eps_psi_approx_J_part} act as fixed peaks, whereas the final Gaussian term is moved to place its peak $x=\mp a$, at such a position which could be used to maximise the integral. We recall from \eqnref{eqn:psi} and \eqnref{eqn:eps} that $\psi$ and $\eps$ contain mixtures of both the real and imaginary part of this integral. When considering the contribution from the imaginary part of the integrand, we see that the maximum of the first complex exponential factor, i.e. of the sinusoidal, will not occur at the origin, unlike that of the second Gaussian factor, and hence in order to maximise this integral comprising three functions we should place the peak of the third function somewhere between the peaks of the first two, which corresponds to a value $a\neq0$. Moreover, as the first sinusoidal factor is periodic, and has a wavelength much smaller than the width $\trm{w}_{\trm{p},0}$, of the Gaussian which multiplies it, we should have a series of maxima in both $\psi(a)$ and $\eps(a)$ which decay slowly with $a$ (see \figref{fig:psi_v_a}). For the case $y_{\trm{d}} \ll y_{\trm{r},\trm{p}}$, our explanation would on the one hand predict that the value of $\eps(a)$ would initially rise as $a$ increases, and on the other hand justify the maximum of $\psi(a)$ being very close to the origin, and hence that $\psi(a)$ would decrease as $a$ initially increases from 0. These results can be further confirmed via differentiation under the integral in \eqnref{eqn:eps_psi_approx}, and are exactly what we observe in the numerical evaluation depicted in \figref{fig:psi_v_a}. This increase is another reflection of the role of Fresnel terms in a non-trivial beam geometry. From numerical analysis, the polarisation and ellipticity were found to increase by a factor of $1.4$ over $a=0$ values.

%

\subsection*{Polarisation results (single-shaft)}
\begin{figure}[!hp]
\noindent
\begin{center}
\includegraphics[width=0.5\linewidth]{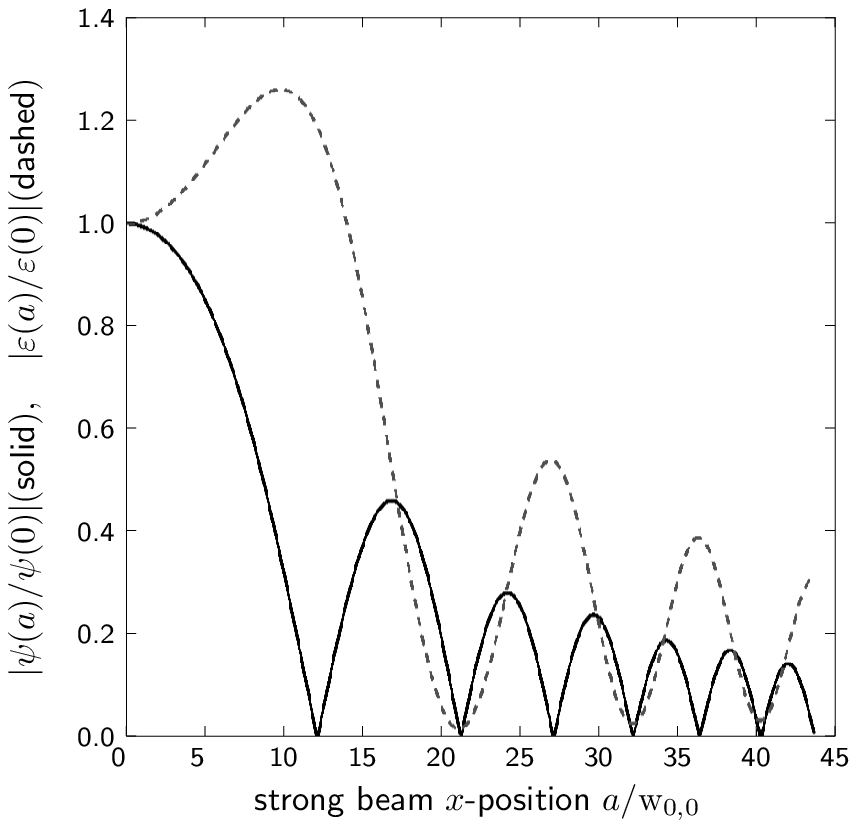}
\end{center}
\caption{The ratios of $|\psi(a) / \psi(0)|$ (continuous line) and $|\eps(a) / \eps(0)|$ (dashed line) in the single-shaft set-up for the same experimental parameters as in Fig. (\ref{fig:phi_diff}) but with $y_{\trm{d}}=50\units{cm}$.\label{fig:psi_v_a_shaft}}
\end{figure}

We want to consider here briefly a different field configuration, in which the strong beams propagate parallel and anti-parallel to the probe field. In this experimental set-up, we simply exchange the co-ordinates $y$ and $z$ in the expressions for the strong fields in Eq. (\ref{eqn:E_0}), gaining a corresponding y-axis strong-field Rayleigh length, $y_{\trm{r},0}$ and Gouy phase $\phi_{\trm{g},0}=\tan^{-1}(y/y_{\trm{r},0})$. To the probe field, we add defocusing terms inside the integral, which would allow us to consider the case $\trm{w}_{\trm{p},0}<\trm{w}_{0,0}$. The diffracted field, $\mbf{E}_{\trm{d}}(\mbfrd)$ is then given by the following expression:
\bea
\mbf{E}_{\trm{d}}(\mbfrd) & := & \frac{I_{0}}{I_{\trm{cr}}} \frac{\alpha E_{\trm{p}}}{45 \lambda^{2}_{\trm{p}} r_{\trm{d}}}
		         \Big( 	2\mathcal{V}'_{1} \mbf{u}'_{1} + 
				(\mathcal{V}'_{3}+\mathcal{V}'_{4}) \mbf{u}'_{34} \Big),\label{eqn:p10}
\eea
\bea \label{eqn:Vk_shaft}
\mathcal{V}'_{\trm{k}} := \!\int^{\infty}_{-\infty} \!\!&d^{3}r\:& \!\!\exp\Big[i\omega_{\trm{p}} \Big(\frac{x^{2} + y^{2} + z^{2}}{2r_{\trm{d}}} - \frac{xx_{\trm{d}}+yy_{\trm{d}}+zz_{\trm{d}}}{r_{\trm{d}}} -\frac{(xx_{\trm{d}} + yy_{\trm{d}} + zz_{\trm{d}})^{2} }{2 r^3_{\trm{d}}} \quad\\
& & +y\Big) - \frac{x^{2} + z^{2}}{\trm{w}_{\trm{p}}^2} - \frac{2}{\trm{w}_{0}^{2}} (x^2+z^2+a^2+b^2) + \frac{i\omega_{\trm{p}}y(x^2+z^2)}{2(y^{2}+y^{2}_{\trm{r},\trm{p}})}\nonumber \\
& &- i\tan^{-1}\Big(\frac{y}{y_{\trm{r},\trm{p}}}\Big)\Big]\nonumber \frac{\mathcal{I}'_{\trm{k}}}{1+(y/y_{\trm{r},0})^2}
\frac{1}{\sqrt{1+(y/y_{\trm{r},\trm{p}})^{2}}} ;
\eea
\bea
\label{eqn:I_1_shaft} \mathcal{I}'_{1} & := & \exp\Big[-\frac{4}{\trm{w}_0^2}\Big( xa + zb\Big)\Big], \nonumber\\
\label{eqn:I_3_shaft} \mathcal{I}'_{3} & := & \exp\Big[\phantom{-}2i\Big(\omega_{0}y - \tan^{-1}\frac{y}{y_{\trm{r},0}} + \frac{\omega_{0}y(x^{2}+z^{2} + a^{2} + b^{2})}{2(y^2 + y_{\trm{r},0}^2)}\Big) +i\Delta\psi_{0}\Big], \nonumber \\
\label{eqn:I_4_shaft} \mathcal{I}'_{4} & := & \exp\Big[-2i\Big(\omega_{0}y - \tan^{-1}\frac{y}{y_{\trm{r},0}} + \frac{\omega_{0}y(x^{2}+z^{2} + a^{2} + b^{2})}{2(y^2 + y_{\trm{r},0}^2)}\Big) -i\Delta\psi_{0}\Big], \nonumber \\
\eea
where we have also introduced a phase difference term $\Delta\psi_0 = \psi_{0,2} - \psi_{0,1}$ between the absolute phases of the two strong beams, which turns out to have negligible effect for the same reasons as separating the beams in the longitudinal direction and is correspondingly set to zero. The vectors $\mbf{u}'_1$ and $\mbf{u}'_{34}$ are:
\bea \label{eqn:I_vecs_1}
\mbf{u}'_{1} & := & \left( \begin{array}{c}
  		(1+\yrd)\cos\theta + \xrd(\xrd\cos\theta + \frac{7}{4}\zrd\sin\theta) \\
  		-\frac{7}{4}\zrd\sin\theta +\yrd\xrd\cos\theta+\frac{7}{4}\xrd\yrd\sin\theta \\
  		\frac{7}{4}\yrd\sin\theta + \xrd\zrd\cos\theta 
			+\frac{7}{4}((\zrd)^{2}+1)\sin\theta
               \end{array} \right), \\ 
\label{eqn:I_vecs_2}
\mbf{u}'_{34} & := & \left( \begin{array}{c} 
		2(1-\yrd)\cos\theta + 2(\xrd)^{2}\cos\theta + -\frac{3}{4}\xrd\zrd\sin\theta \\
		3\xrd\cos\theta - \frac{3}{4}\zrd\sin\theta(\yrd+1)+2\xrd\yrd\cos\theta \\
                \frac{3}{4}(\yrd-1)\sin\theta + 2\xrd\zrd\cos\theta - \frac{3}{4}(\zrd)^{2}\sin\theta
                          \end{array} \right).
\eea
As in the previous case, we set $(x_{\trm{d}}/r_{\trm{d}}), (z_{\trm{d}}/r_{\trm{d}}) \rightarrow 0, (y_{\trm{d}}/r_{\trm{d}}) \rightarrow 1$ in \eqnref{eqn:I_vecs_1} and \eqnref{eqn:I_vecs_2} which removes the latter vector completely, eliminating any  contribution from $\mbf{E}_{0,2}(\mathbf{r},t)$, the strong-field beam with wavevector parallel to that of the probe. This result is expected from the general property of a plane wave that it does not polarise the vacuum. In this geometry, we obtain for the polarisation $\psi$ and the ellipticity $\eps$ the following expressions:
\protect\bea \label{eqn:psi2}
\psi = \frac{\alpha\sin2\theta}{15\lambda_{\trm{p}}^{2}}\frac{I_{0}}{I_{\trm{cr}}}
\Big( \frac{\mathcal{V}_{1}^{\prime r}}{y_{\trm{d}}}
+\frac{\mathcal{V}_{1}^{\prime i}}{y_{\trm{r},\trm{p}}}\Big),\\
\label{eqn:eps2}
\eps = \frac{\alpha\sin2\theta}{15\lambda_{\trm{p}}^{2}}\frac{I_{0}}{I_{\trm{cr}}}
\Big(\frac{\mathcal{V}_{1}^{\prime r}}{y_{\trm{r},\trm{p}}}
-\frac{\mathcal{V}_{1}^{\prime i}}{y_{\trm{d}}}\Big).
\eea

We can compare these to existing results arrived at by Heinzl et al. \cite{heinzl_birefringence06} when we take $a=b=0$ and the two limits: the \emph{refractive-index} and the \emph{crossed-field} limit. The first is obtained when we take $y_{\trm{d}}\rightarrow0$ (near region), in a regime where $\psi$ becomes linear with $y_{\trm{d}}$ and therefore disappears, and $\eps$ converges to a constant. The crossed-field limit corresponds to a constant strong field, i.e. $\omega_{0}\rightarrow0$, which we can achieve when we let the counter-propagating pulse be e.g. of the form of a cosine. This ensures that neither the strong electric nor magnetic field disappears in this limit, so that we can keep the normalisation used in \eqnrefs{eqn:psi2}{eqn:eps2}. To be consistent, the time-averaging procedure which removes evanescent waves must be repeated with the precondition that $\omega_{0}=0$. Then \eqnref{eqn:eps2} tends to the result in \cite{heinzl_birefringence06}:
\bea
\eps = \frac{2\alpha\pi}{15}\frac{I_{0}}{I_{\trm{cr}}}\frac{y_{0}}{\lambda_{\trm{p}}}\sin2\theta; \qquad\qquad
y_{0} = \frac{y_{\trm{r},\trm{p}}y_{\trm{r},0}}{y_{\trm{r},\trm{p}}+y_{\trm{r},0}}.
\eea
The only difference to the formula in \cite{heinzl_birefringence06} is that we have incorporated the focusing of the strong- and probe- fields, which automatically generates the effective interaction length $y_{0}$ of the beams.
\newline

Another feature which is different here, is that we allow the strong-field wave to be positioned off-axis. We showed and explained how this increases the ellipticity and polarisation in the double-slit set-up, and in this single-shaft experiment with just one beam, one acquires a similar result (see \figref{fig:psi_v_a_shaft}). For the same experimental parameters as in Fig. (\ref{fig:phi_diff}) but with $y_{\trm{d}}=50\units{cm}$ and $a/\trm{w}_{0,0}=10$, we achieve a modest increase in the ellipticity of $1.3$ over single strong-beam values.  We mention here, that one could also form a double-shaft geometry which leads to the same relative increase as for the off-axis single-shaft one. As $x_{\trm{d}}=z_{\trm{d}}=0$, this can be understood as a result of the symmetry of the set-up, which can also be seen in \eqnref{eqn:Vk_shaft}, being symmetric in $x$ and $a$ in this limit.

\section{Conclusion}
\setcounter{equation}{0}
\renewcommand{\theequation}{\Roman{section}.\arabic{equation}}
A main focus of this paper was to extend the results derived in \cite{dipiazza_PRL_06} to incorporate more features applicable to experiment. One development has been to extend into the far-field region, the range in which polarisation rotation and ellipticity formulae are valid. These results were calculated for two different geometries: double-slit and single/double-shaft. Another addition has been to include a separation of the strong-field beams. This non-trivial beam geometry in conjunction with higher-order Fresnel diffraction terms was shown to increase polarisation rotation and ellipticity values for a range of beam parameters in the double-slit case by a factor of $1.4$, and in the single/double-shaft case by a factor of $1.3$. Although these increases are relative to the values at zero beam separation, we acknowledge that since the overall accuracy of the calculation is $\approx1/\pi$, some care should be taken in interpreting these results. By calculating the diffraction pattern resulting from the interference between the polarised vacuum and probe signals, we have illuminated another possible route to measuring laser-induced VPEs. For experimental parameters comfortably attainable at the upcoming X-FEL and ELI facilities, we have shown how approximately $10^{-5}$ of the incident photons can be diffracted, with around two photons per shot of the lasers being diffracted into regions where the vacuum signal is higher than the probe background. However, we stress that an increase would also be observed using ELI with a table-top X-ray laser such as e.g. in \cite{kapteyn99} where a beam of frequency $29~\trm{nm}$ was used. These, in principle measurable diffraction vacuum polarisation effects, would be the first evidence of non-linear vacuum polarisation in laser fields.
\newline

\noindent{\large \bf{Acknowledgement} }\\
We would like to acknowledge helpful discussions with J. Crespo Lopez-Urrutia.
\clearpage

\section*{Appendix}\label{app:A}
\setcounter{equation}{0}
\renewcommand{\theequation}{A.\arabic{equation}}
The volume integral from \eqnref{eqn:Vk} can be integrated in the $x$ and $y$ co-ordinates to give:

\bea 
\mathcal{V}_{\trm{k}} = \!\int^{\infty}_{-\infty} \!\!&dz&\: \frac{\pi \trm{w}^{2}_{0,0}}{\sqrt{\alpha_{x} \alpha_{y}}}\frac{1}{1+(z/z_{\trm{r},0})^{2}}\exp\Big\{\frac{\pi^{2}}{\alpha_{y}}
\left(\frac{\trm{w}_{0,0}}{\lambda_{\trm{p}}} \right)^{2} \Big[i\frac{y_{\trm{d}}}{r_{\trm{d}}}\Big(1+ \frac{zz_{\trm{d}}}{r^{2}_{\trm{d}}}\Big) \label{eqn:p12-in-1}\\ 
& & +~ i\frac{\pi}{\alpha_{x}}\frac{x_{\trm{d}}y_{\trm{d}}}{r_{\trm{d}}^{2}}\frac{\trm{w}_{0}^{2}}{r_{\trm{d}}\lambda_{\trm{p}}} \Big(\frac{i x_{\trm{d}}}{r_{\trm{d}}} + \frac{i x_{\trm{d}}z_{\trm{d}}z}{r_{\trm{d}}^3} - \frac{\beta_{\trm{k}}}{\pi}\frac{a \lambda_{\trm{p}}}{\trm{w}_{0}^{2}}\Big) -\frac{\beta_{\trm{k}}}{\pi}\frac{b \lambda_{\trm{p}}}{\trm{w}_{0}^{2}} -i\Big]^{2} \nonumber\\
& & -~ \frac{z^{2}}{\trm{w}_{\trm{p},0}^{2}}\Big[ \frac{i \pi \trm{w}_{\trm{p},0}^{2}}{\lambda_{\trm{p}} r_{\trm{d}}}\Big( 1-\Big(\frac{z_{\trm{d}}}{r_{\trm{d}}}\Big)^{2}\Big) + 1 \Big] + \nonumber \frac{4\pi}{\alpha_{x}}\left(\frac{\trm{w}_{0,0}}{\lambda_{\trm{p}}}\right)^{2} \Big[i\frac{x_{\trm{d}}}{r_{\trm{d}}} \Big(1+ \frac{zz_{\trm{d}}}{r_{\trm{d}}^{2}}\Big) - \frac{\beta_{\trm{k}}}{\pi}\frac{a \lambda_{\trm{p}}}{\trm{w}_{0}^{2}}\Big]^{2} \\
& & +~ 2\pi i \frac{z_{\trm{d}}}{r_{\trm{d}}}\frac{z}{\lambda_{\trm{p}}} + 4i\Gamma_{\trm{k}} \pi \frac{z}{\lambda_{0}} \Big[ 1 + \frac{a^{2} + b^{2}}{2(z^{2} + z_{\trm{r},0}^{2})}\Big] \nonumber - 2i\Gamma_{\trm{k}}\phi_{\trm{g},0}(z) - \frac{2(a^{2} + b^{2})}{\trm{w}_{0}^{2}}\Big\}, \nonumber
\eea
\newline
where we have defined:
\bea 
\alpha_{x} \!& := &\! i \pi \frac{\trm{w}_{0,0}^{2}}{\lambda_{\trm{p}}r_{\trm{d}}}\Bigg[1\!-\!\left(\!\frac{x_{\trm{d}}}{r_{\trm{d}}}\!\right)^{2}\!\Bigg] + 
\frac{2}{1+(z/z_{\trm{r},0})^{2}} - 
\frac{2i\Gamma_{\trm{k}} z}{z_{\trm{r},0}}\frac{1}{1 + (z/z_{\trm{r},0})^{2}} + \left(\frac{\trm{w}_{0,0}}{\trm{w}_{\trm{p},0}}\right)^{2}\label{eqn:alpha_x},\\
\alpha_{y} \!& := &\! i \pi \frac{\trm{w}_{0,0}^{2}}{\lambda_{\trm{p}}r_{\trm{d}}}\Bigg[1\!-\!\left(\!\frac{y_{\trm{d}}}{r_{\trm{d}}}\!\right)^{2}\!\Bigg] + 
\frac{2}{1+(z/z_{\trm{r},0})^{2}} - 
\frac{2i\Gamma_{\trm{k}} z}{z_{\trm{r},0}}\frac{1}{1 + (z/z_{\trm{r},0})^{2}} + \frac{\pi^{2}}{\alpha_{x}}\left(\frac{x_{\trm{d}}y_{\trm{d}}}{r_{\trm{d}}^{2}}\frac{\trm{w}_{0}^{2}}{r_{\trm{d}}\lambda_{\trm{p}}}\right)^{2}\!\!,\;\label{eqn:alpha_y}\quad
\eea
\newline
and included all four integrals with:
\be
\Gamma_{\trm{k}} = \left\{ \\
\begin{array}{rl}
1 & \trm{if } k=1, \\
-1 & \trm{if } k=2, \\
0 & \trm{if } k=3,4,
\end{array}  \right.
\qquad \trm{ and } \qquad
\beta_{\trm{k}} = \left\{ \\
\begin{array}{rl}
0 & \trm{if } k=1,2, \\
1 & \trm{if } k=3, \\
-1 & \trm{if } k=4.
\end{array}  \right.
\ee

\bibliography{standard}
\end{document}